\documentclass[a4paper,fleqn,usenatbib]{mnras}

\usepackage{newtxtext,newtxmath}
\usepackage[T1]{fontenc}
\usepackage{ae,aecompl}

\usepackage{graphicx}	% Including figure files
\title[Multiwavelength Variability in V404~Cyg]{Optical and X-ray Correlations During the
2015 Outburst \\of the Black Hole V404~Cyg}

\author[R.~I. Hynes et al.]{R.~I. Hynes$^{1}$\thanks{E-mail: rih@phys.lsu.edu},
E.~L. Robinson$^{2}$, D.~M. Terndrup$^{3}$, P. Gandhi$^{4}$, C.~S. Froning$^{2}$, \newauthor R.~M.
Wagner$^{3,5}$, S. Starrfield$^{6}$, V.~S. Dhillon$^{7,8}$, T.~R. Marsh$^{9}$\\
$^{1}$Department of Physics and Astronomy, Louisiana State
University, Baton Rouge, Louisiana 70803, USA\\
$^{2}$Department of Astronomy, University of Texas at Austin, 1
University Station, Austin, TX 78712, USA\\
$^{3}$Department of Astronomy, The Ohio State University, 140 W. 18th
Avenue, Columbus, OH 43210, USA\\
$^{4}$Department of Physics and Astronomy, University of Southampton, Highfield, Southampton SO17 1BJ\\
$^{5}$Large Binocular Telescope Observatory, 933 North Cherry Avenue,
Tucson, AZ 85721, USA\\
$^{6}$School of Earth and Space Exploration, Arizona State University,
Tempe, AZ 85287-1404, USA\\
$^{7}$Department of Physics and Astronomy, University of Sheffield, Sheffield S3 7RH, UK\\
$^{8}$Instituto de Astrof\'{i}sica de Canarias, E-38205 La Laguna, Tenerife, Spain\\
$^{9}$Department of Physics, University of Warwick, Coventry CV4 7AL, UK}

\date{Accepted XXX. Received YYY; in original form ZZZ}

\pubyear{2018}

\begin{document}
\label{firstpage}
\pagerange{\pageref{firstpage}--\pageref{lastpage}}
\maketitle

% Abstract of the paper Limit to 250 words
\begin{abstract}
  We present a serendipitous multiwavelength campaign of optical
  photometry simultaneous with {\it Integral} X-ray monitoring of the
  2015 outburst of the black hole V404~Cyg. Large amplitude optical
  variability is generally correlated with X-rays, with lags of order
  a minute or less compatible with binary light travel timescales or
  jet ejections. Rapid optical flaring on time-scales of seconds or
  less is incompatible with binary light-travel timescales and has
  instead been associated with synchrotron emission from a jet. Both
  this rapid jet response and the lagged and smeared one can be
  present simultaneously.  The optical brightness is not uniquely
  determined by the X-ray brightness, but the X-ray/optical
  relationship is bounded by a lower-envelope such that at any given
  optical brightness there is a maximum X-ray brightness seen.
  This lower-envelope traces out a
  $F_{\rm opt} \propto F_{\rm X}^{0.54}$ relation which can be
  approximately extrapolated back to quiescence.  Rapid optical
  variability is only seen near this envelope, and these periods
  correspond to the hardest hard X-ray colours. This correlation
  between hard X-ray colour and optical variability (and
  anti-correlation with optical brightness) is a novel finding of this
  campaign, and apparently a facet of the outburst behaviour in
  V404~Cyg. It is likely that these correlations are driven by changes in the
central accretion rate and geometry.
\end{abstract}


%%%%%%%%%%%%%%%%%%%%%%%%%%%%%%%%%%%%%%%%%%%%%%%%%%

%%%%%%%%%%%%%%%%% BODY OF PAPER %%%%%%%%%%%%%%%%%%

\section{Introduction}
\label{IntroSection}

Some low-mass X-ray binaries (LMXBs) undergo dramatic X-ray, optical,
and radio outbursts, separated by years, decades, or even longer
periods of quiescence \citep{McClintock:2006a}.  These transient LMXBs
have a much higher incidence of black holes than persistent systems
\citep{King:1996a,King:1997a,King:1997b}, hence they are often known
as Black hole X-ray transients (BHXRTs).  The class is diverse, with
orbital periods of hours to days and companion stars of spectral types
B--M including main-sequence stars, sub-giants and giants.  While some
objects follow a common fast-rise, exponential-decay (FRED) pattern,
many do not, and outburst morphology is varied
\citep{Chen:1997a,Tetarenko:2016a}. A subset of BHXRTs are known as
microquasars due to the presence of resolved, relativistic
jets. Possibly most sources produce jets which are either too compact
to resolve, or not observed at the right time.

Among both BHXRTs and neutron star LMXBs distinct states have been
identified \citep{Remillard:2006a}. These correspond to different
modes of accretion, with different geometrical
configurations. Multiple states may be possible at the same accretion
rate. Historically X-ray colour-colour and hardness-intensity diagrams
have played a major role in classifying these states. Over the last
fifteen years diagrams initially of X-ray luminosity vs. radio
luminosity \citep[e.g.][]{Corbel:2003a,Gallo:2003a}, and later
infrared, or optical luminosity
\citep[e.g.][]{Homan:2005a,Russell:2006a} have been added to the
repertoire of diagnostic tools, and have provided valuable insights
into the relationship between the canonical states and jet
production. In the optical, in particular, \citet{Russell:2006a}
demonstrated a clear correlation between X-ray and optical
luminosities in the hard spectral state with
$L_{\rm opt} \propto L_{\rm X}^{0.61}$, which extended from outburst
to quiescence. In principle both X-ray reprocessing in the accretion
disc, and direct jet emission could give rise to this correlation. The
IR, and in some cases the optical flux are suppressed in the
thermal-dominated spectral state, at just the time that radio emission
is suppressed, suggesting that jet emission is the dominant mechanism
driving the correlation, at least some of the time. This is supported
by additional evidence for strong, rapid optical/IR variability in the
hard state which cannot be explained by disc reprocessing
\citep{Motch:1982a,Motch:1983a,Kanbach:2001a,Spruit:2002a,
  Hynes:2003a,Hynes:2006a,Durant:2008a,Gandhi:2008a,Durant:2009a,
  Hynes:2009a,Gandhi:2010a,Durant:2011a}, highlighting the way for
detailed, high time-resolution multiwavelength observations to
disentangle the roles of disc-reprocessing and jet emission in driving
the correlation.

The X-ray nova GS~2023+332 was discovered with {\it Ginga} on 1989 May
22 as a very hard and highly variable X-ray source
\citep{Makino:1989a}.  Several groups independently realised that the
optical source coincided with Nova Cyg 1938 = V404~Cyg
\citep{Wagner:1989a,Wagner:1991a}.  When V404~Cyg returned to
quiescence it was possible to study the system dynamically and
elucidate system parameters.  It is among the longest period BHXRTs
with a 6.5~day orbital period \citep{Casares:1992a}. The black hole
mass is estimated at $9.0^{+0.2}_{-0.6}$~M$_{\odot}$
\citep{Khargharia:2010a}.  It is the most X-ray luminous BHXRT in
quiescence with the best studied high energy properties in the
class. Quiescent X-ray observations have found a range of X-ray
luminosities of $4\times10^{32} - 3\times10^{33}$~erg\,s$^{-1}$ (see
compilation in \citealt{Bradley:2007a} and
\citealt{Bernardini:2014a}), with larger amplitude variability seen
within long observations
\citep[e.g.][]{Hynes:2004a,Bradley:2007a,Rana:2015a}.  Simultaneous
multiwavelength observations have shown that both optical lines and
continuum correlate clearly with X-rays, with optical line emission,
at least, appearing to arise from disc reprocessing
\citep{Hynes:2004a}.  X-ray and optical emission are correlated with
$L_{\rm opt} \propto L_{\rm X}^{0.56}$ in outburst and a slightly
flatter relation $L_{\rm opt} \propto L_{\rm X}^{0.46}$ in quiescence
\citep{Bernardini:2016a}.  Radio emission, while securely detected and
quite variable, is not clearly correlated with X-rays within an
observation \citep{Hynes:2009b,Rana:2015a}, in spite of the longterm
correlation \citep{Gallo:2003a}.  Very Long Baseline Interferometer
radio observations have also permitted measurement of a parallax
distance of $2.39\pm0.14$~kpc \citep{MillerJones:2009a} facilitating
estimates of multiwavelength luminosity, albeit subject to
uncertainties in the local absorption and emission geometry.

On 2015 June 15 a new outburst was heralded by gamma-ray burst
triggers on {\it Swift} \citep{Barthelmy:2015a}, {\it Fermi}
\citep{Younes:2015a}, and {\it Konus-Wind} \citep{Golenetskii:2015a}.
Once the X-ray outburst itself begin, an extensive {\it INTEGRAL}
monitoring campaign
\citep{Rodriguez:2015a,Natalucci:2015a,Roques:2015a,Siegert:2016a,SanchezFernandez:2017a,Jourdain:2017a,Motta:2017a,Rodi:2017a},
complemented by observations by {\it Swift}
\citep{Radhika:2016a,Motta:2017a,Motta:2017b}, {\it Fermi}
\citep{Jenke:2016a,Loh:2016a}, {\it NuStar} \citep{Walton:2017a}, and
{\it Agile} \citep{Piano:2017a} revealed that the high energy
behaviour was dominated by flares showing up to a three order of
magnitude increase in brightness over the inter-flare brightness,
peaking around the Eddington limit. The X-ray spectrum showed complex,
multi-component variations. The underlying hard X-ray spectrum could
be modelled with two components. \citet{SanchezFernandez:2017a}
classified the hard spectral states into two branches: a hard branch
that appears comparable to the canonical hard state in other BHXRTs,
and a soft flaring branch that is similar to canonical intermediate
states. No true thermal dominated state is seen. A low-energy thermal
Comptonization component dominates the continuum.  A high-energy
power-law was also seen possibly also related to synchrotron jet
emission, or non-thermal Comptonization
\citep{Rodriguez:2015a,Roques:2015a}. During flares, a large
relativistic reflection component was seen indicating a disc extending
close to the last stable orbit and likely a rapidly spinning black
hole \citep{Walton:2017a}. At low energies some evidence was suggested
for direct emission from a disc blackbody component
\citep{Radhika:2016a,Walton:2017a}, although \citet{Motta:2017b}
notably could not confirm this with reanalysis of the {\it Swift}
dataset.  Transient absorption was probably present most of the time,
with very high absorption columns, and appears to consist of two
components, an extended homogeneous low density component, and a
compact clumpy component reaching very high column densities
\citep{Motta:2017b}.  Material above the disc, sometimes associated
with an outflow, was indicated by a rich X-ray emission line spectrum
seen at high resolutions \citep{King:2015b}. Evidence for a strong
wind was also seen in optical spectroscopic monitoring
\citep{MunozDarias:2016a}. This appeared to carry away a large amount
of mass, shutting off the outburst and resulting in a brief nebular
phase \citep{MunozDarias:2016a,Rahoui:2017a}.  \citet{Kimura:2016a}
examined an extensive database of optical monitoring throughout the
outburst, providing an invaluable context to studies based on more
limited coverage. They found that the optical was already bright
within minutes of the first trigger by {\it Swift} and showed dramatic
variability throughout the outburst, characterized by episodes of both
flaring and dipping. Generally the optical was found to be correlated
with X-rays.  The outburst was quite short-lived, fading rapidly in
2015 July and reaching apparent X-ray quiescence by July 23, less than
40~days after the outburst began \citep{Plotkin:2017a}.

We report here on the properties of optical variability on
timescales of seconds and longer, and their relationship to the X-ray
and gamma-ray variability seen by {\it Integral}.  Preliminary
analyses of the optical data presented here were reported by
\citet{Hynes:2015a,Hynes:2015b,Terndrup:2015a,Gandhi:2015a,Gandhi:2015b}. A
detailed study of the short timescale Ultracam data have been reported
separately by \citet{Gandhi:2016a} and \citet{Gandhi:2017a}.

\section{Observations}

\subsection{McDonald Observatory}

We observed V404~Cyg on the nights beginning 2015 June 17 and 18 using
the Argos CCD Photometer on the 2.1~m telescope at McDonald
Observatory. Details of the observations are given in
Table~\ref{LogTable}. Observing conditions were good with mostly
stable transparency and typical seeing of 1~arcsec on both nights.

\begin{table}
\centering
\caption{Log of optical time-series observations}
\label{LogTable}
\begin{tabular}{llcc}
\hline
Date & Observatory & Filter & Time \\
(UTC) &                    &          & Resolution\\
\hline
June 18.27--18.44 & McDonald & $r'$ & 2~s\\
June 19.35--19.44 & McDonald & $r'$ & 1~s\\
\noalign{\smallskip}
June 19.21--19.48 & MDM & $R$ & 6~s\\
June 20.20--20.47 & MDM & $R$ & 1~s\\
June 21.20--21.46 & MDM & $R$ & 1~s\\
June 22.19--22.44 & MDM & $R$ & 1~s\\
\noalign{\smallskip}     
June 20.18--20.23 & WHT & $r'$ & 1~s\\
June 21.16--21.22 & WHT & $r'$ & 1~s\\
June 25.15--25.23 & WHT & $r'$ & 1~s\\
June 26.21--26.21 & WHT & $r'$ & 1~s\\
June 26.21--26.22 & WHT & $r'$ & 1~s\\
\hline
\end{tabular}
\end{table}

On each night we obtained bias frames, dark frames, and dome
flat-fields. The images were processed in the usual way using {\sc
  iraf}\footnote{IRAF is distributed by the National Optical Astronomy
  Observatory, which is operated by the Association of Universities
  for Research in Astronomy (AURA) under a cooperative agreement with
  the National Science Foundation.}.  V404~Cyg was the brightest star
in the field by a large margin, so we did not perform simple
differential photometry. We extracted aperture photometry of both
V404~Cyg, and the second brightest star in the field,
IPHAS2~J202403.00+335129.3
\citep[$r'=15.98$][]{Drew:2005a,Barentsen:2014a}, using an 8~pixel
(3~arcsec) aperture. The aperture was chosen to include the light from
the faint contaminating star 1.4~arcsec north of V404~Cyg. The
comparison star was used to establish a flux calibration, estimate an
extinction correction, and to correct for slow transparency
variations. Since the comparison star was fainter than V404~Cyg, we
estimated the gradual transparency variations by fitting a spline
function to its lightcurve after correcting for extinction.

The contaminating star has magnitude $V=18.90\pm0.02$,
$R=17.52\pm0.01$ \citep{Casares:1993a}. Using the transformation
equations from \citet{Jordi:2006a} this corresponds to
$r'=18.21\pm0.07$.  We can also see this star directly when V404~Cyg
is faintest. Using the ten images with best seeing around the period
of minimum light on the first night we construct an average image and
measure the brightness of the contaminating star relative to our
comparison star using {\sc daophot} in {\sc iraf}, leading to an
estimate of $r=18.11\pm0.06$. These two methods give quite consistent
estimates, so we use their average, $r=18.16$, to subtract the light
from the contaminating star from V404~Cyg.

All Argos exposures are synced to begin on GPS ticks, and measured
relative to a known coordinated universal time (UTC) start time. For
comparison with other datasets we converted the lightcurves to
heliocentric Julian dates, HJD$_{\rm UTC}$ at mid-exposure.

\subsection{MDM Observatory}

We also observed V404~Cyg on the nights beginning 2015 June 18--21
using the Andor frame-transfer CCD on the 1.3~m McGraw Hill Telescope
at MDM Observatory.  A brief description of these data was presented
in \citet{Terndrup:2015a}.  Details of the observations are given in
Table~\ref{LogTable}.  The observations were obtained in the R-band at
an image scale of 0.54~arcsec/pixel and with a field-of-view of
2.3~arcmin. The imaging on June 19 consisted of 1.29~h at a
cadence of 10.0~s, followed by 2.10 and 2.94~h at a cadence
of 7.5 and 6.0~s, respectively.  The photometry on the remaining
nights was obtained with a 1~s cadence.  All times were converted
to HJD$_{\rm UTC}$ at mid-exposure.

We removed a baseline from each frame using an average of many
zero-second exposures, and flattened them using twilight sky images.
Conditions were mostly non-photometric, so we report differential
photometry of V404 Cyg with respect to the anonymous field star
2MASS~J20240718+3350516 (AUID~000-BCL-467 in the AAVSO field
photometry database and 620--101865 in the UCAC4 catalog). This star
has $R=12.384$ mag based on photometry obtained by A. Henden and
reported in the AAVSO Variable Star
Database\footnote{http://www.aavso.org/apps/vsp} allowing calibration into
$R$ magnitudes.  We obtained aperture photometry using a radius equal
to 2.5 times each image's FWHM; the latter was taken as the average
value for V404~Cyg and the comparison star.  Sky levels were computed
using a median value surrounding each star. Errors in the photometry
were estimated by the distribution of differences from each frame to
the next and how these were correlated with the brightness of V404
Cyg.  Typical errors in the differential photometry are about 0.02
mag.  Light from the nearby contaminating star was subtracted assuming
it has $R=17.52\pm0.01$ \citep{Casares:1993a}.

\subsection{William Herschel Telescope}

Finally, shorter duration but very high time-resolution observations
were obtained using Ultracam on the 4.2~m William Herschel Telescope
(WHT) on the nights beginning 2015 June 19, 20, 24, and 25. Partial
overlaps were obtained with MDM coverage. Details of the observations
are given in Table~\ref{LogTable}.  The data reduction procedures for
Ultracam data have been described in detail by
\citet{Gandhi:2016a}. For this work we use only 1~s time
resolution lightcurves in the $r'$ filter, converted to
HJD$_{\rm UTC}$ at mid-exposure. Light from the nearby contaminating
star was subtracted in the same way as for the McDonald data.

\subsection{Las Cumbres Observatory}

Additional non-time-series monitoring was performed on four nights
using the Las Cumbres Observatory (LCO; \citealt{Brown:2013a}),
summarized in Table~\ref{LCOTable}. These extend our coverage into
quiescence. We observed V404 Cyg in the $ugri$ filters on each night,
varying the exposure times based on the expected brightness of the
target but typically observing in total for 10--45~min in each
filter. Some of the data were not usable because of exposure time
mis-estimates for the auto-scheduled observations: the $u$ and $g$
observations on 30 June and the $u$ observations on 12 July could not
be used because we underexposed while the $i$ data on 26 June could
not be used because the individual exposures were saturated on V404
Cyg.

The data were downloaded from the LCO archive, which provides images
calibrated using the LCO pipeline. Aperture photometry was performed
for V404 Cyg and several field stars using IRAF. For the 24 and 26
June observations we extracted each frame individually and propagated
the scatter about the mean through the flux calibration
calculation. For 30 June and 12 July, when V404 Cyg had faded, we
coadded the individual exposures before extraction. We extracted
several field stars in addition to V404 Cyg in each frame and
calculated relative photometry for each date. As V404 Cyg became
fainter, we changed the comparison stars to fainter ones to maintain
comparable brightnesses. The data were flux calibrated using 24 June
observations of a nearby standard stars (Wolf 1346). For the $r$ and
$i$ data in 30 June and 12 July we improved the flux calibration by
adding four nearby field stars from the IPHAS Data Release 2
\citep{Barentsen:2014a} to the relative photometry calculations. We
did not correct the photometry in Table~\ref{LCOTable} for
contribution of the nearby contaminating star, although we did correct
$r$ magnitudes for inclusion in Fig.~\ref{LongLCFig}.

\begin{table}
\centering
\caption{LCO Photometry}
\label{LCOTable}
\begin{tabular}{lllc}
\hline
Date & Filter & MJD & Mag. \\
(UTC) &          & (Mid)          \\
\hline
June 24 & $u$ & 57198.28832 & $14.17\pm0.09$ \\
& $g$ & 57198.29957 & $12.36\pm0.11$ \\
& $r$ &  57198.30438 & $10.81\pm0.03$ \\
& $i$ & 57198.30942 & $9.54\pm0.15$ \\
\noalign{\smallskip}     
June 26 & $u$ & 57199.37727 & $13.98\pm0.18$ \\
& $g$ & 57199.34991 & $12.69\pm0.06$\\
& $r$ & 57199.35764 & $11.16\pm0.05$\\
\noalign{\smallskip}     
June 30 & $r$ & 57203.34508 & $15.72\pm0.09$ \\
& $i$ & 57203.35176 & $14.18\pm0.09$ \\
\noalign{\smallskip}     
July 12 & $g$ & 57215.38353 &$18.93\pm0.09$ \\
& $r$ & 57215.39157 & $17.08\pm0.05$\\
& $i$ & 57215.39717 & $14.97\pm0.07$\\
\hline
\end{tabular}
\end{table}

\subsection{Integral}
\label{IntegralSection}

V404~Cyg was observed extensively by {\em Integral} throughout the
2015 outburst. Public data products have been made available
\citep{Kuulkers:2015a} and are used for comparison with optical
lightcurves in this work. We also extracted higher time-resolution
lightcurves from the raw data.

JEM-X processed lightcurves were provided at 8~s time-resolution
separately for the JEM-X1 and JEM-X2 cameras, divided into 5--10~keV
and 10-25~keV bands. We combined summed data from both cameras and
both channels to create a single 5--25~keV X-ray lightcurve with
8~s time-resolution, as well as combining the two channels to
produce a soft colour index ($[10-25{\rm keV}] / [5-10{\rm
  keV}]$).
ISGRI lightcurves come from a single camera and were provided in
25--60~keV and 60--200~keV energy bands at 64~s
time-resolution. Again we combined these into a single 25--200~keV
gamma-ray lightcurve, and also constructed a hard colour index
($[60-200{\rm keV}] / [25-60{\rm keV}]$). Finally we combined the
instruments to produce an overall colour index
($[25-200{\rm keV}] / [5-25{\rm keV}]$).

To work in approximate fluxes and luminosities, we converted count rates
for each of the four {\it Integral} bands defined above into fluxes using
conversions derived using {\sc webpimms}, assuming power-law spectra
with photon indices $\Gamma = 1.0, 1.5, 2.5, 3.0$ for 5--10, 10--25,
25--60, and 60--200~keV respectively. The photon indices were chosen
to approximately match the shape of the spectra in each band shown by
\citet{Rodriguez:2015a}. A constant interstellar absorption column of
$N_{\rm H} = 0.88\times10^{22}$~cm$^{-2}$ \citep{Bradley:2007a} was
assumed to estimate unabsorbed fluxes. These were then summed and
converted to an isotropically emitted luminosity assuming a distance
of 2.39~kpc \citep{MillerJones:2009a}. Note that this remains subject
to uncertainties about emission geometry and variable local absorption and
cannot be considered equivalent to an inferred accretion rate. In
particular, we have only corrected for interstellar absorption, and
flux variations due to local absorption will remain. 
The inferred luminosities (5-200~keV) during the periods of
simultaneous coverage range from
$2\times10^{35}$~erg~s$^{-1}$ to $6\times10^{38}$~erg~s$^{-1}$, peaking
at $\sim50$~percent of the Eddington limit for a 9~M$_{\odot}$ black
hole.  At the highest luminosities the ISGRI band still contains more
flux than JEM-X indicating a quite hard state.

Finally, for cross-correlation against rapid optical lightcurves, we
construct a JEM-X lightcurve at 1~s time-resolution for a single
5--25~keV bandpass. For this the original data were extracted from 
HEASARC, and processed using {\sc osa} v10.2 to produce LCR
lightcurves. 

All {\it Integral} times are provided in ISDC Julian Date (IJD) format
in the Terrestrial Time (TT) system, i.e. IJD$_{\rm TT}$. We convert
from TT first to International Atomic Time (TAI) by subtracting
31.184~s, then to UTC by subtracting a further 35 leap seconds. We
converted IJD to HJD by adding 2,451.544.5 and applying a heliocentric
correction. Our final times are thus in the HJD$_{\rm UTC}$ system
consistent with our optical data.

\subsection{Cross calibration of optical datasets}

\begin{figure*}
\includegraphics[width=7.2in, trim=0.7in 0.0in 0.0in 0.0in, clip]{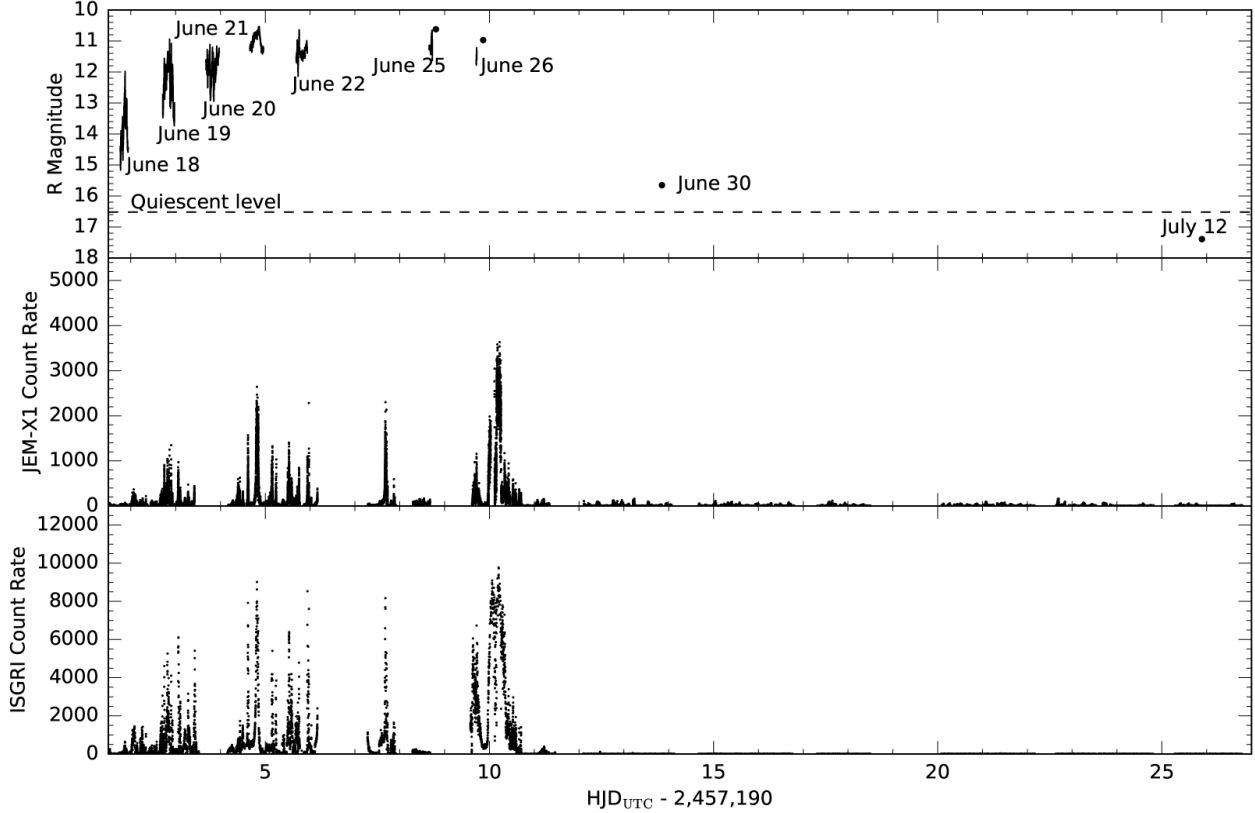}
\caption{Overall outburst lightcurves based on time-resolved
  data. {\em Integral} data are shown with 8~s (JEM-X) and 64~s
  (ISGRI) resolution to preserve the extremely transient nature of the
detected X-ray activity.}
\label{LongLCFig}
\end{figure*}

We were fortunate in obtaining overlaps between both McDonald and MDM,
and the WHT and MDM, in spite of the serendipitous nature of our
observations. We can use these to verify both the relative timing and
photometric calibration, and in some cases the reality of unusual
features of the lightcurves.

Both Argos and Ultracam are synced to GPS time, and should be reliable
to much better than the 1~s time resolution employed in this
work. To verify the timing calibration of the MDM data, we
cross-correlate overlapping lightcurves against both Argos and
Ultracam. On June 19 we find $t_{\rm McD} - t_{\rm MDM}=4.5$~s,
with MDM data taken at 6~s time resolution. On June 20 we find
$t_{\rm WHT} - t_{\rm MDM} = 2.3$~s, with 1~s resolution,
and on June 24 we find $t_{\rm WHT} - t_{\rm MDM} = -1.5$~s,
with 1~s time resolution but very limited overlapping data. We
conclude that timing errors in the MDM data are at most a few
seconds. They are certainly smaller than the 8~s time-resolution
of the public JEM-X data products, but may lead to small offsets with
respect to 1~s JEM-X lightcurves.

We find small differences in the flux calibration between overlapping
datasets. This is not surprising as MDM used the $R$ filter
($\lambda_{\rm eff} = 6588$\ \AA) rather than $r'$
($\lambda_{\rm eff} = 6290$\ \AA). We measure $r'-R=0.15$ between
McDonald and MDM on June 19, and $r'-R=0.23$ between WHT and MDM on
both June 20 and June 24. We attribute these differences primarily to
the difference between $r'$ and $R$ filters, and to a lesser extent
the crude calibration applied, neglecting colour terms between the
target and comparison stars. Where appropriate we convert both
McDonald and WHT data to the $R$ band for comparison using the colour
terms measured from simultaneous data. For LCO $r$ data we adopt
$r-R=0.19$ as the average of McDonald and WHT measurements.

\section{Lightcurves}
\label{Lightcurve Section}

\subsection{Overall Morphology and Slow Variations}
\label{MorphologySection}

\begin{figure*}
\includegraphics[width=3.4in]{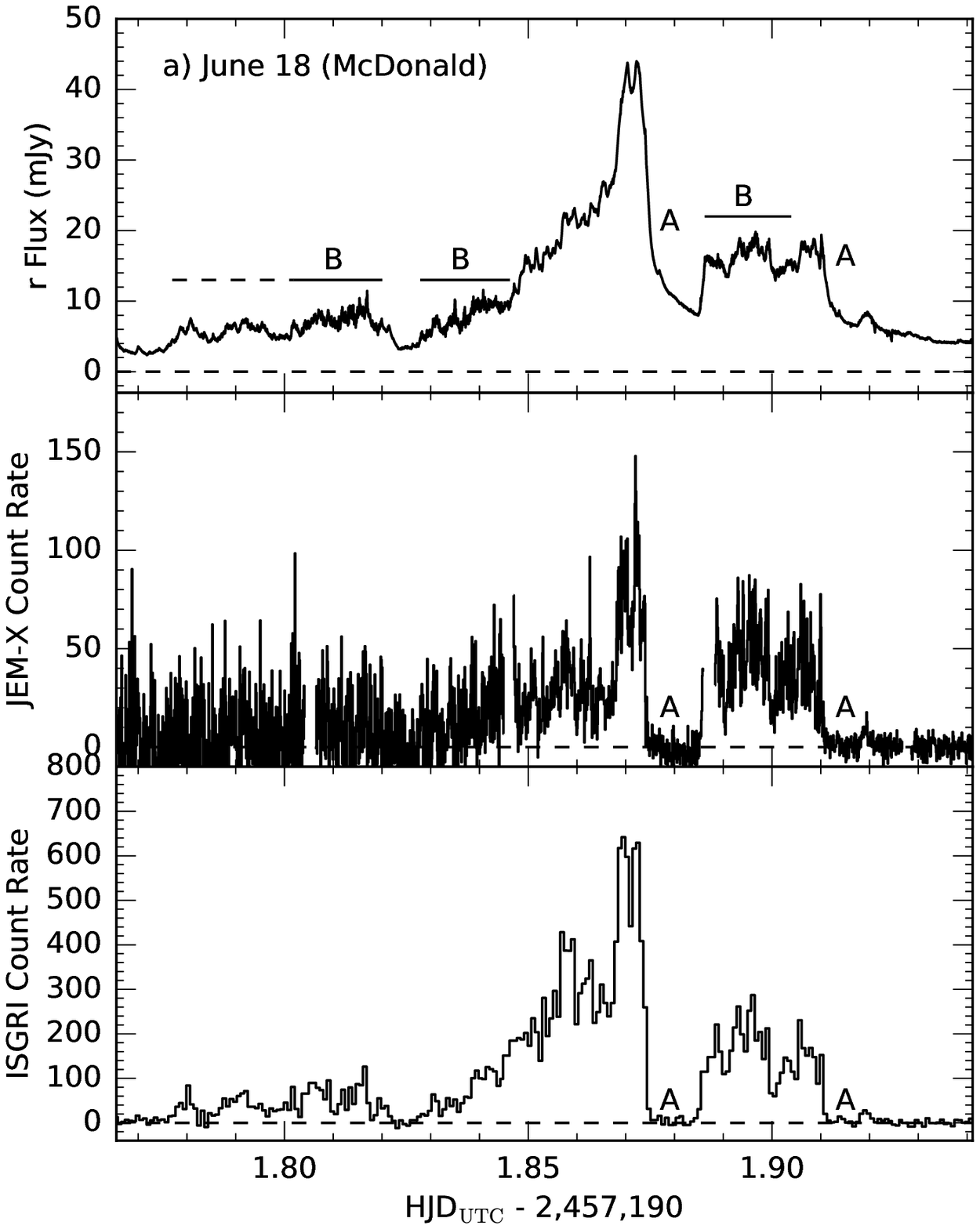}
\hspace*{\fill}
\includegraphics[width=3.4in]{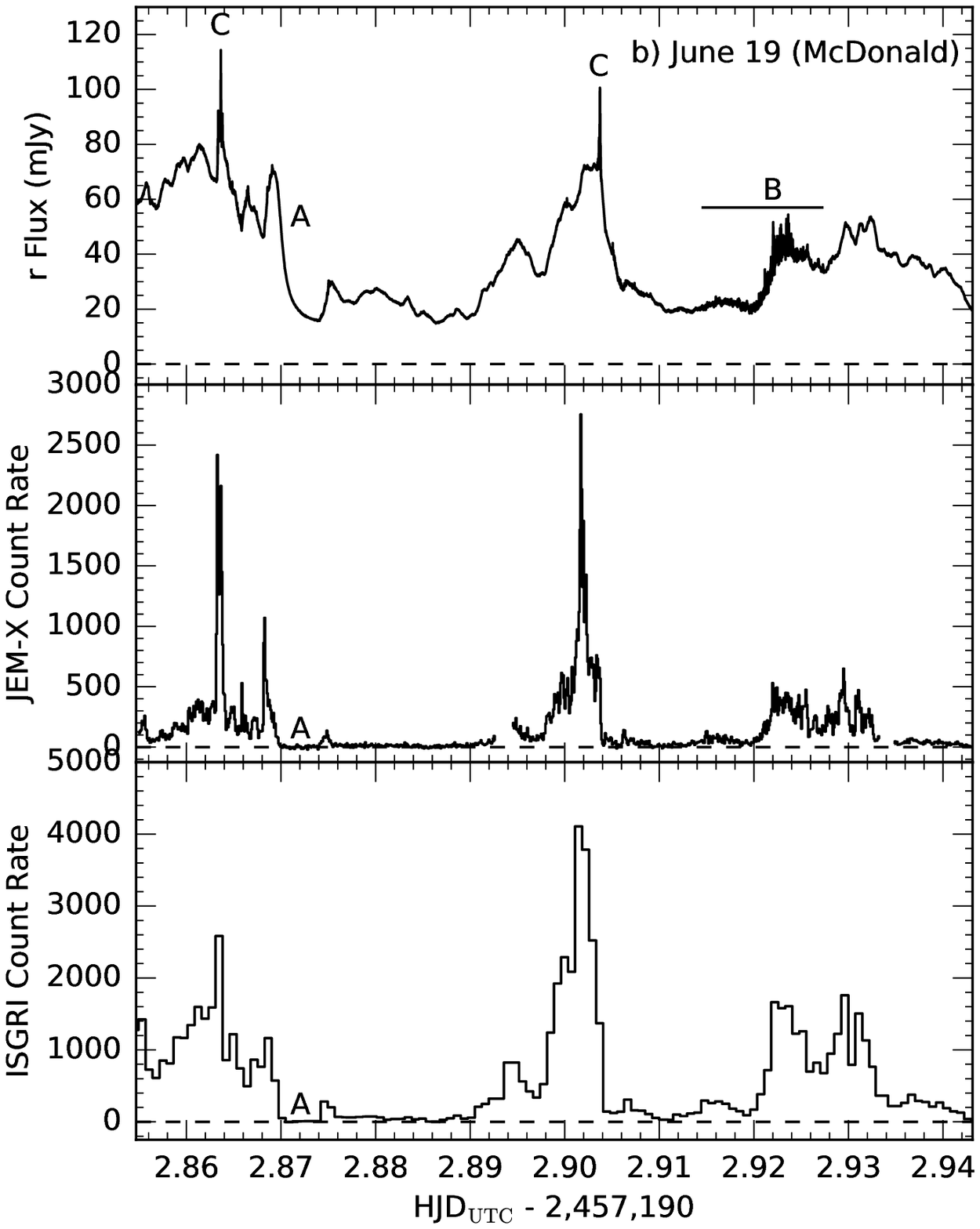}
\caption{McDonald data compared to Integral
  data. Labels A denote smooth decays, B indicate
 periods of rapid optical flaring, and C denotes fast flares at
  peak light.}
\label{McDFig}
\end{figure*}

\begin{figure*}
\includegraphics[width=3.4in, trim=0.0in 0.0in 0.0in 0.0in, clip]{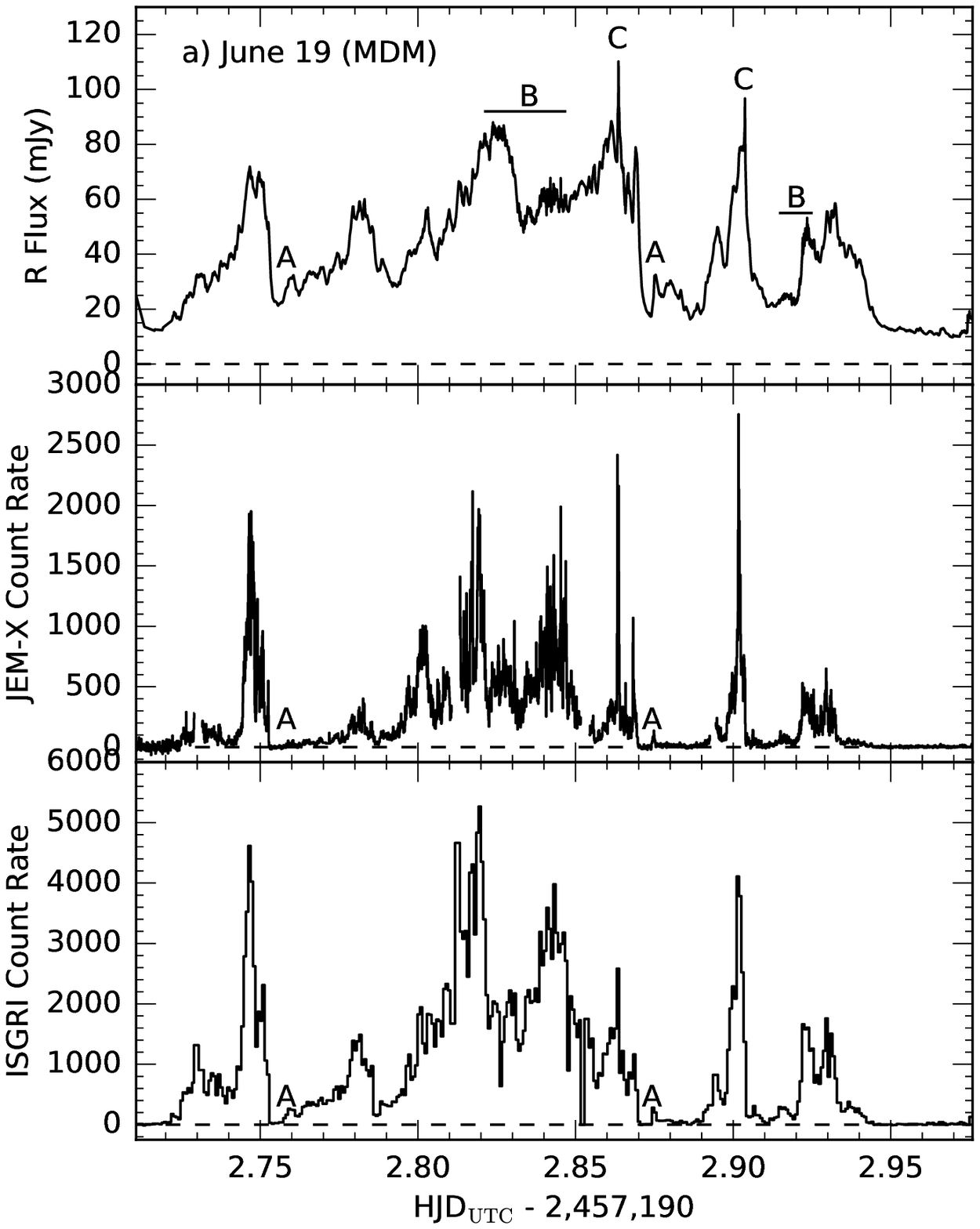}
\hspace*{\fill}
\includegraphics[width=3.4in, trim=0.0in 0.0in 0.0in 0.0in, clip]{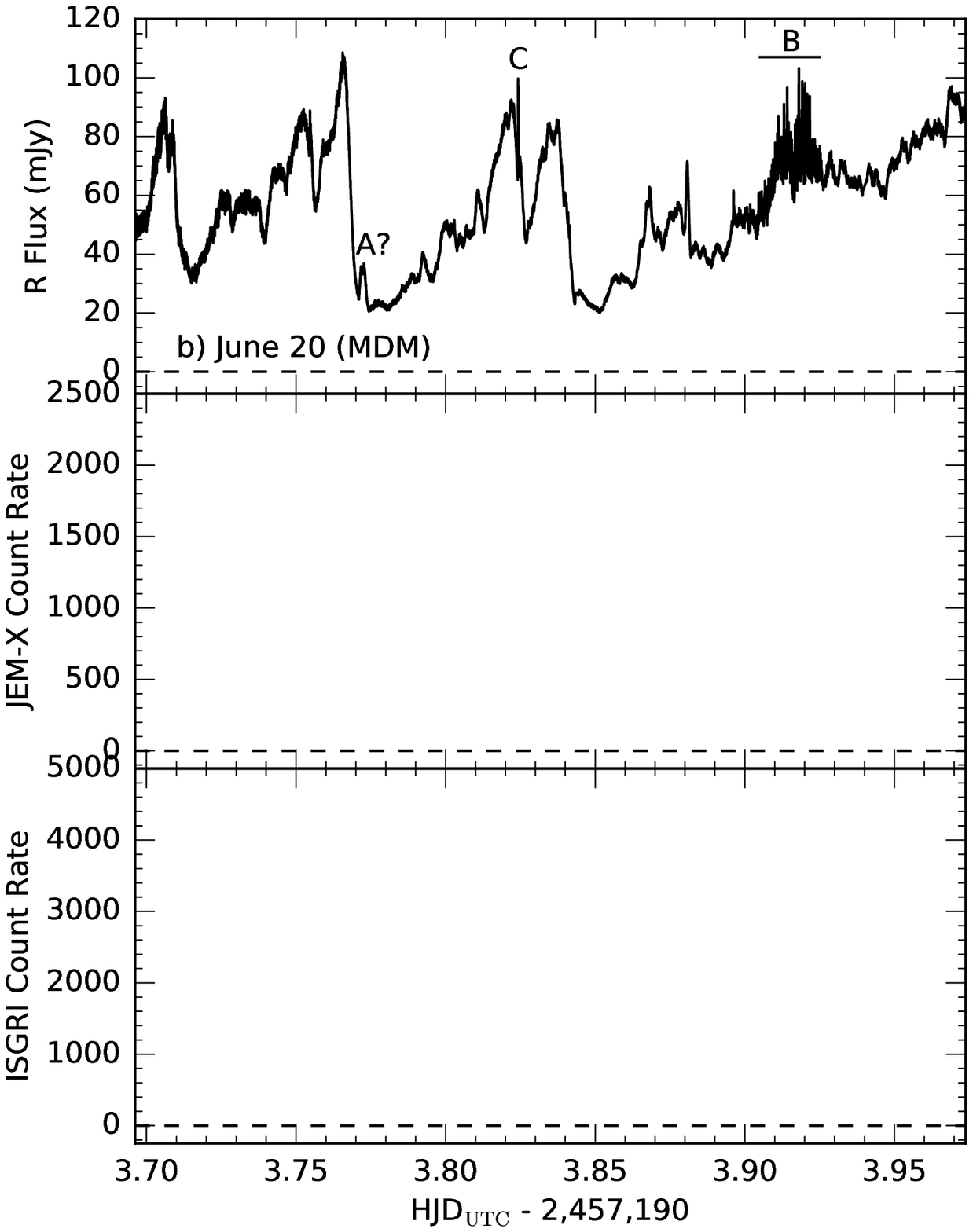}\\
\includegraphics[width=3.4in, trim=0.0in 0.0in 0.0in 0.0in, clip]{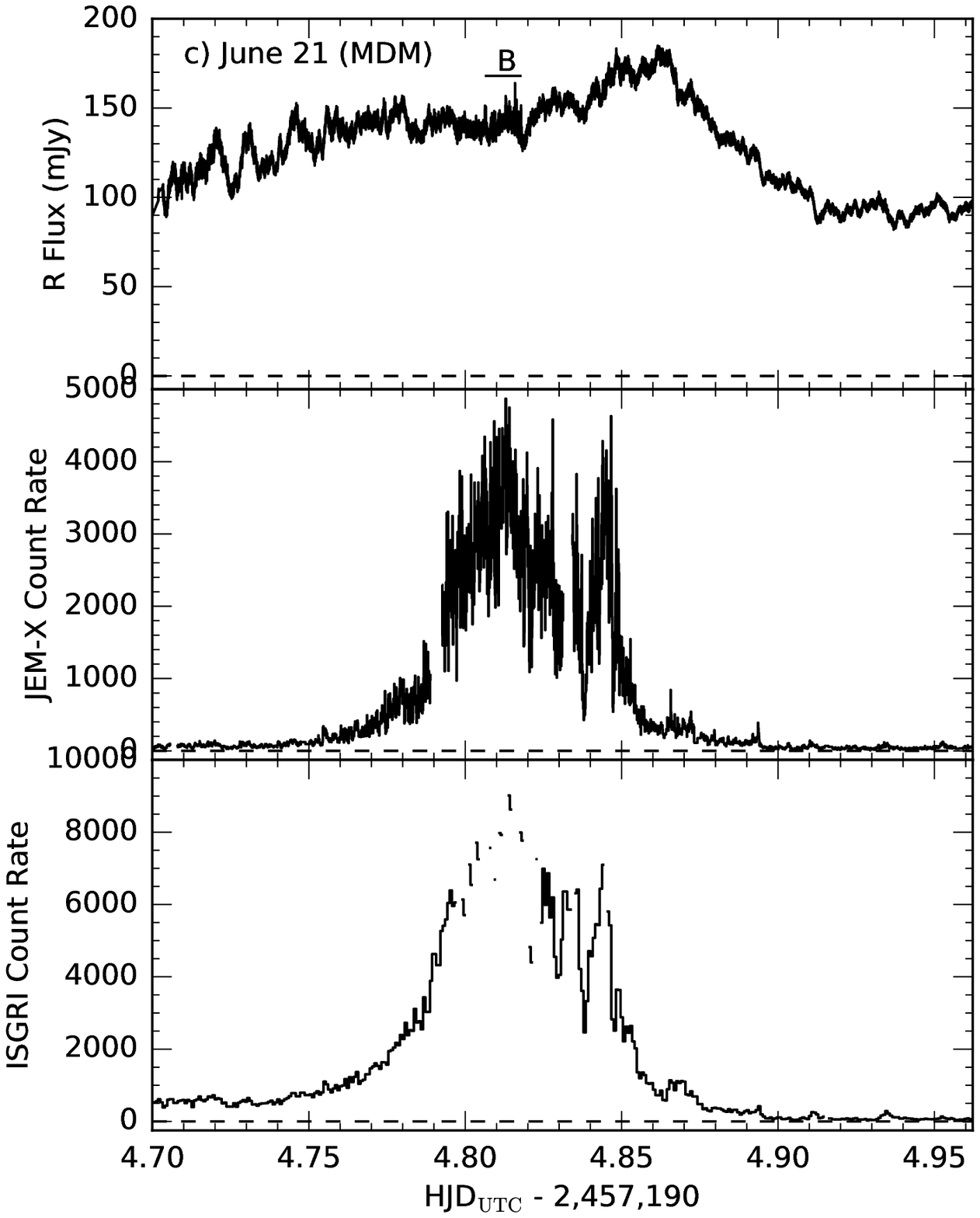}
\hspace*{\fill}
\includegraphics[width=3.4in, trim=0.0in 0.0in 0.0in 0.0in, clip]{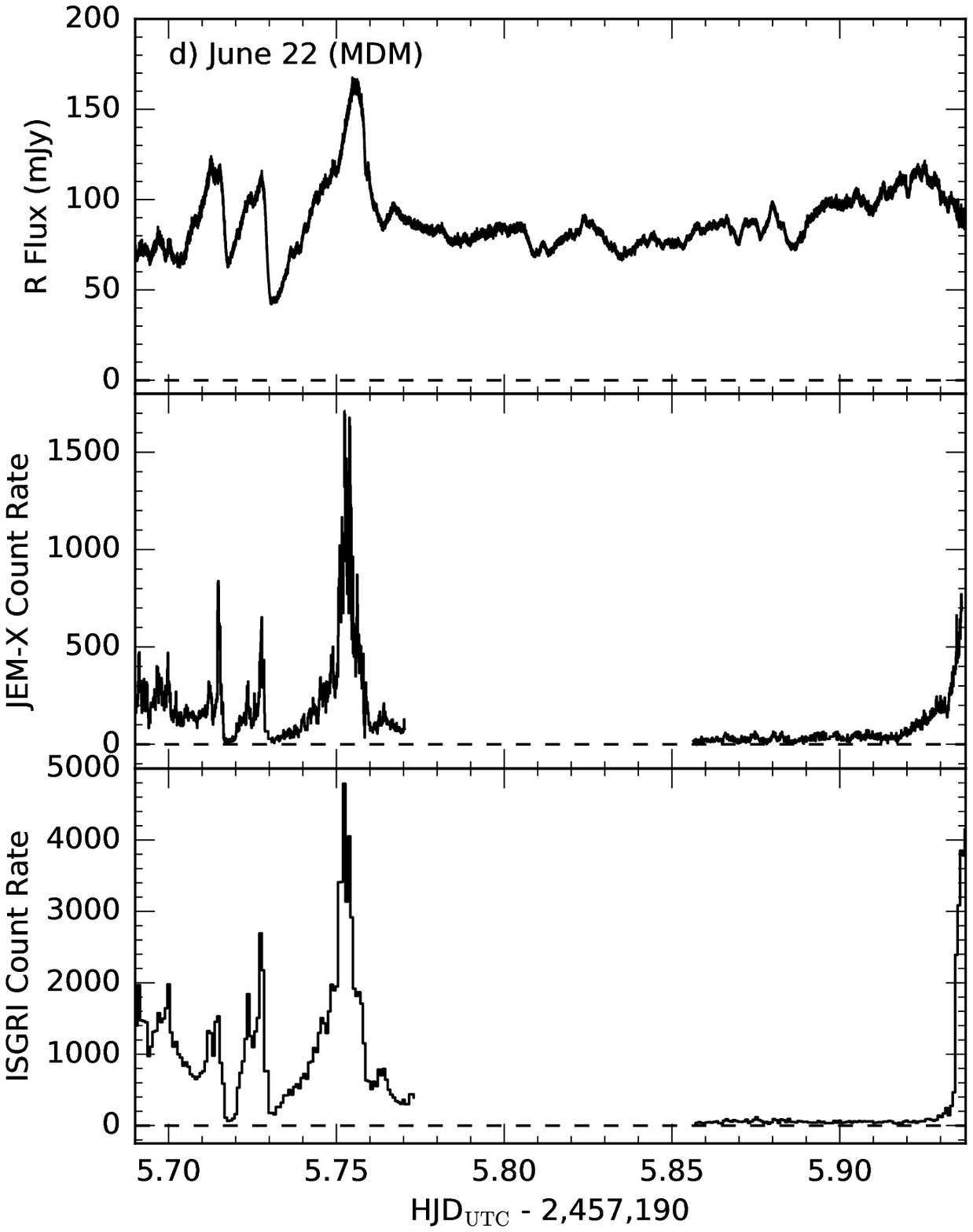}
\caption{MDM data compared to {\it Integral} data. Annotations are the same
  as in Fig.~\ref{McDFig}. Where {\it Integral} panels are blank, no
  data were taken at that time.}
\label{MDMFig}
\end{figure*}

\begin{figure*}
\includegraphics[width=3.4in, trim=0.0in 0.0in 0.0in 0.0in, clip]{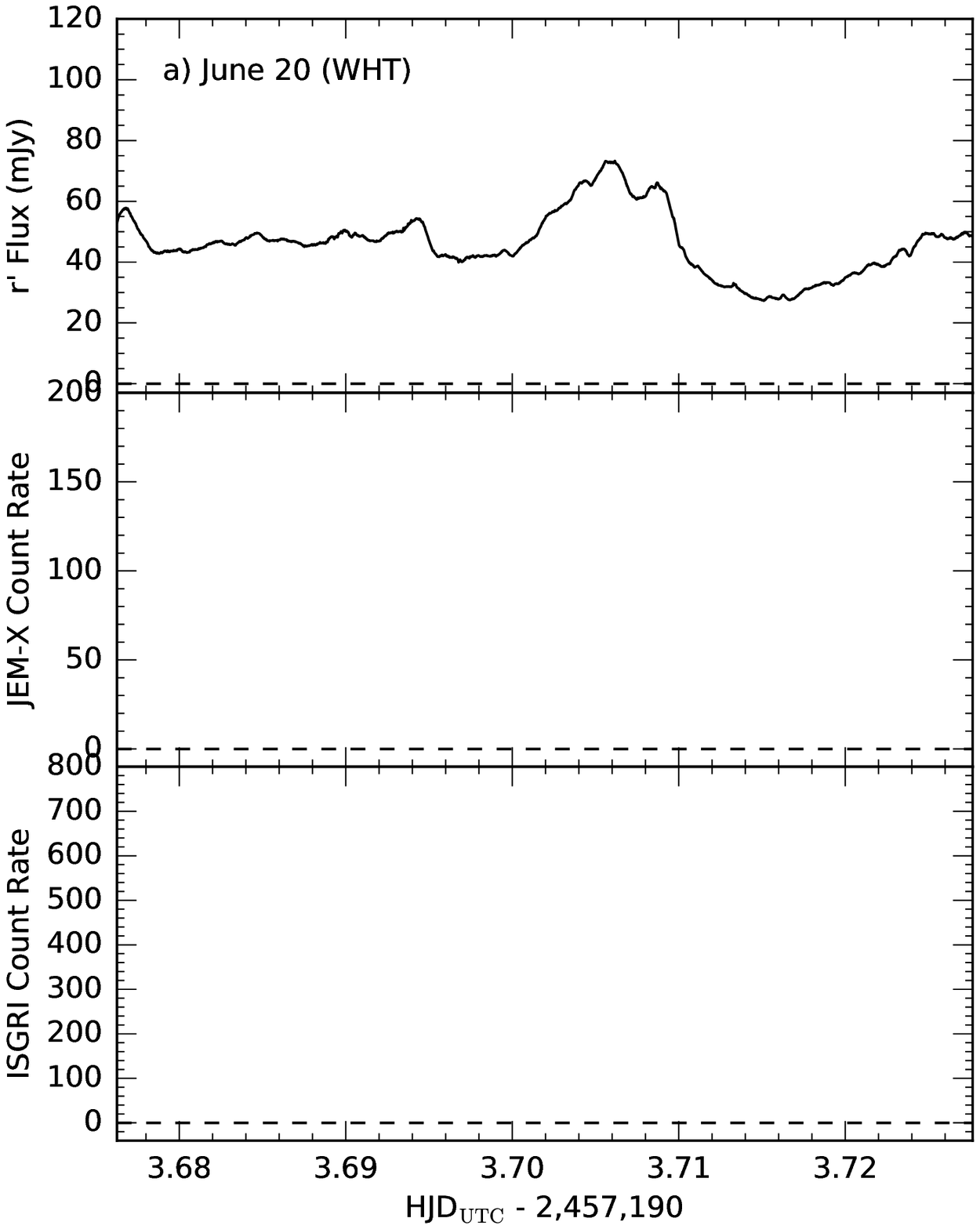}
\hspace*{\fill}
\includegraphics[width=3.4in, trim=0.0in 0.0in 0.0in 0.0in, clip]{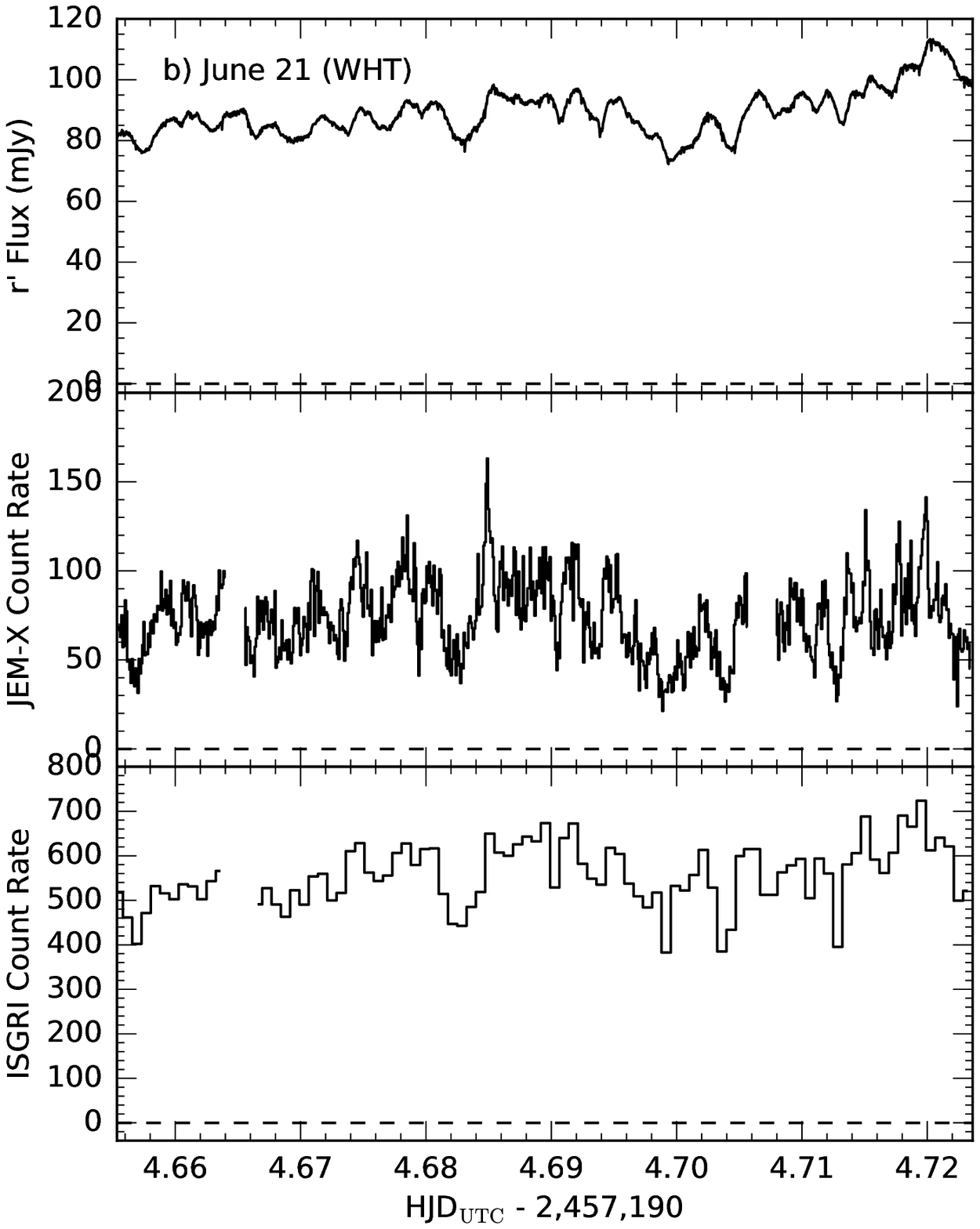}\\
\includegraphics[width=3.4in, trim=0.0in 0.0in 0.0in 0.0in, clip]{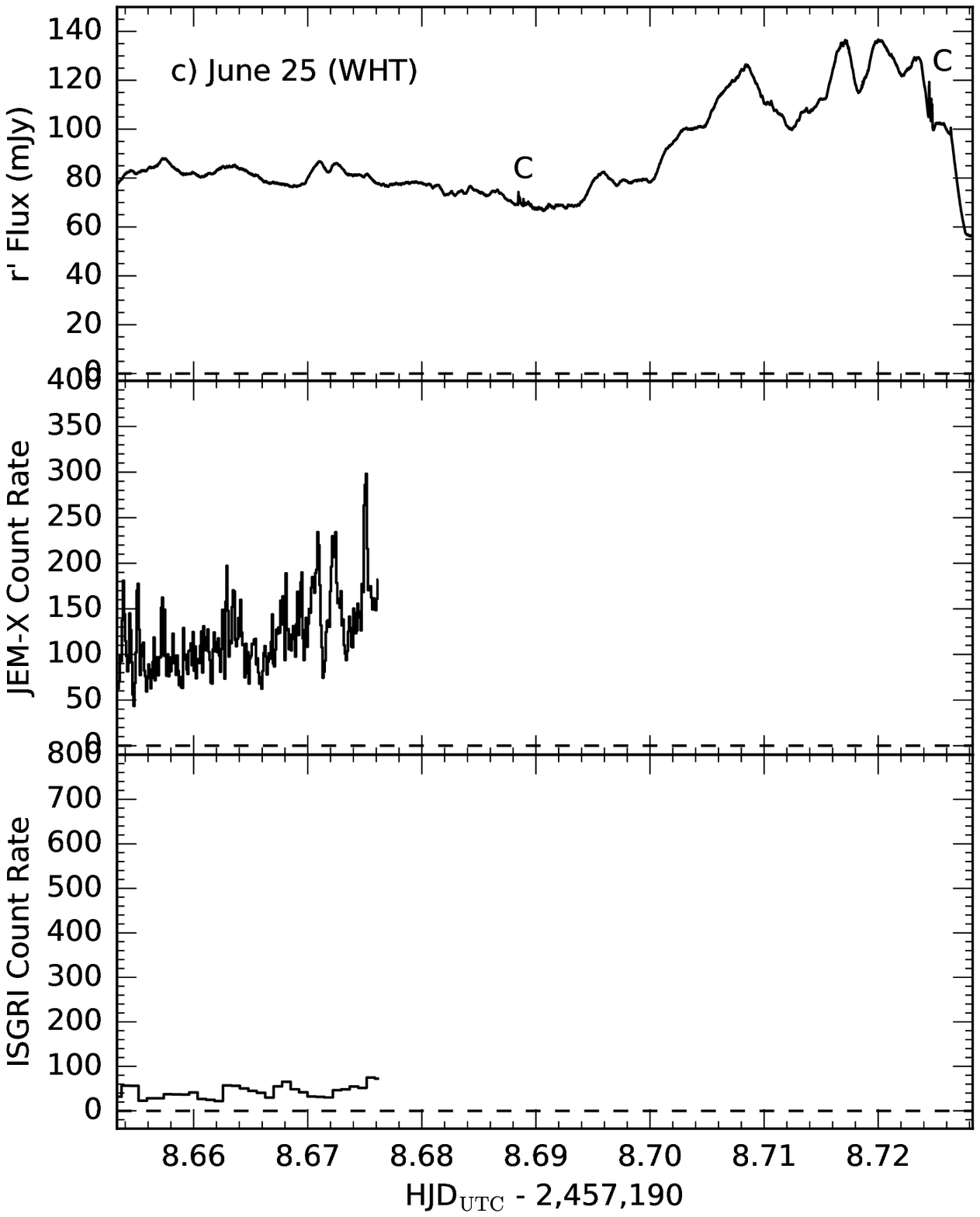}
\hspace*{\fill}
\includegraphics[width=3.4in, trim=0.0in 0.0in 0.0in 0.0in, clip]{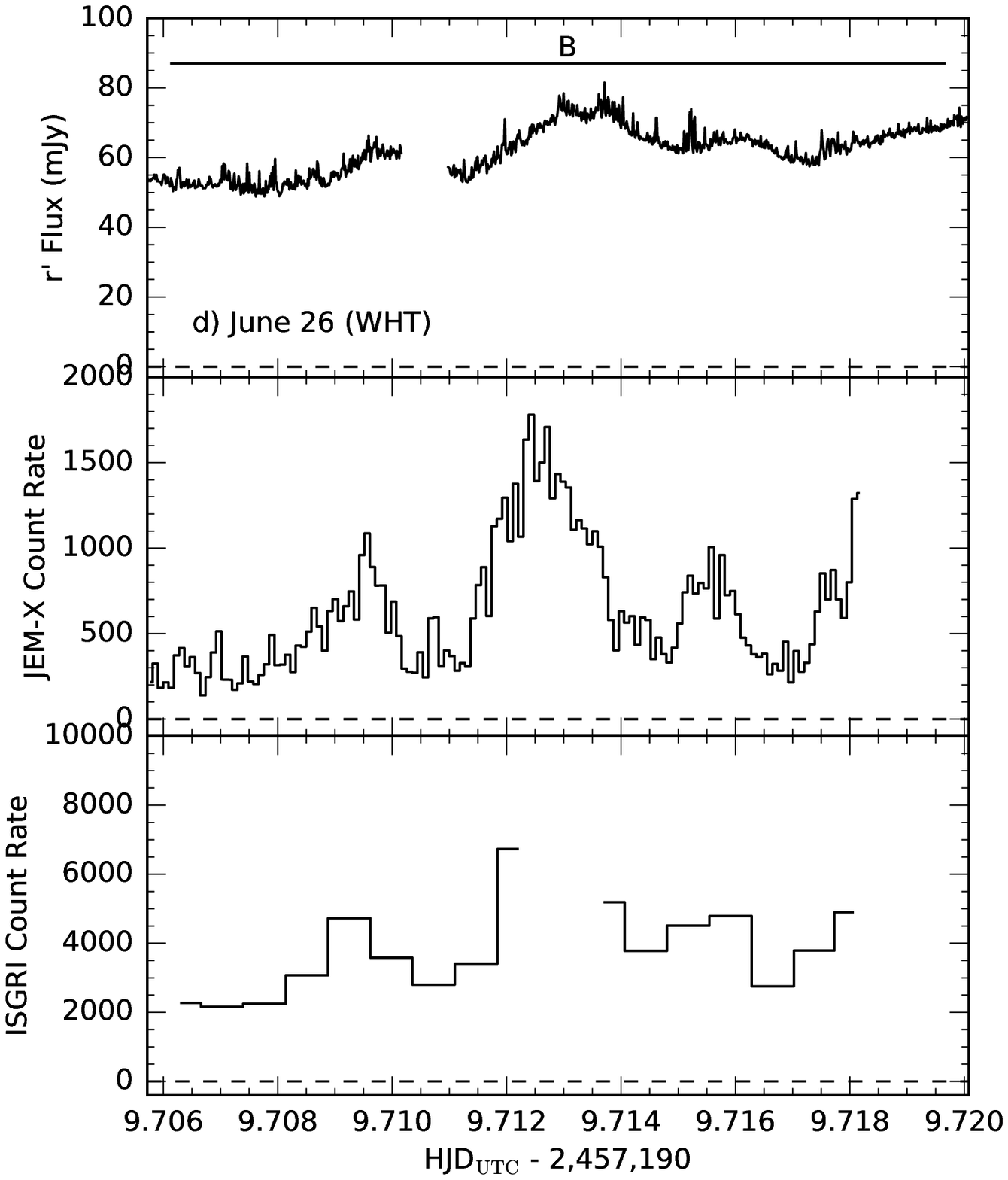}
\caption{WHT/Ultracam data compared to {\it Integral} data. Annotations are the same
  as in Fig.~\ref{McDFig}. Where {\it Integral} panels are blank, no
  data were taken at that time.}
\label{UltracamFig}
\end{figure*}

We compile all the optical lightcurves, transformed to $R$, in
Fig.~\ref{LongLCFig}, and show individual nights' lightcurves
converted to fluxes in Figs.~\ref{McDFig}--\ref{UltracamFig}.  We
began our coverage when the source was hitting minimum brightness
about a magnitude above quiescence, and followed through the rise to
an apparent plateau and then the drop back to quiescence. Our coverage
is much less intensively sampled than that of \citet{Kimura:2016a}
which reveals a more complex behaviour. The rising phase involved
continual large amplitude variability, with minimum and maximum
  magnitudes generally rising. Following \citet{Gandhi:2016a}, we
refer to all of this heterogeneous behaviour which is ubiquitously
present to some degree in most lightcurves on timescales of minutes or
longer, as slow variations.  At times shorter timescale flares are
superposed on the slow variations
(Section~\ref{OpticalTransitionSection}). The plateau phase exhibited
a generally more stable level interspersed with dips or recurrent
oscillations. Shortly after the our last time resolved observation,
\citet{Kimura:2016a} saw the source decline back to fainter levels, as
also seen in our LCO data.

\citet{MunozDarias:2016a} show that H$\alpha$ equivalent widths were
below $\sim300$~\AA\ until June 28, so in all of our time-resolved
observations, the $r'$ and $R$ band \citep[FWHM$\sim1400$~\AA\ and
1600~\AA\ respectively;][]{Fukugita:1995a} should be dominated by
continuum light. While the line contribution is not negligible, the
large amplitude variations seen cannot originate in emission lines
alone.

We see dramatic optical variations during the rise. In this period
there are multiple occasions when the X-ray flux vanishes (at the
limits of {\em Integral} sensitivity), and the optical flux is
minimized. Generally, a correspondence can be seen between features in
X-ray and optical lightcurves, although this is not perfect in detail,
and virtually breaks down on June 21.  Nonetheless, the similarities
between the detailed X-ray and optical lightcurves are more prominent
than their differences), and this is most clearly apparent at high
energies. The JEM-X lightcurve, which is likely strongly affected by
absorption \citep{Motta:2017a,Motta:2017b}, is notably more transient
than the higher energy ISGRI one, and quite a good correspondence is
in fact present between optical and ISGRI lightcurves in
Figs.~\ref{McDFig}--\ref{UltracamFig}.

\subsection{Smooth Decays}
\label{DecaySection}

\begin{figure*}
\includegraphics[width=3.4in]{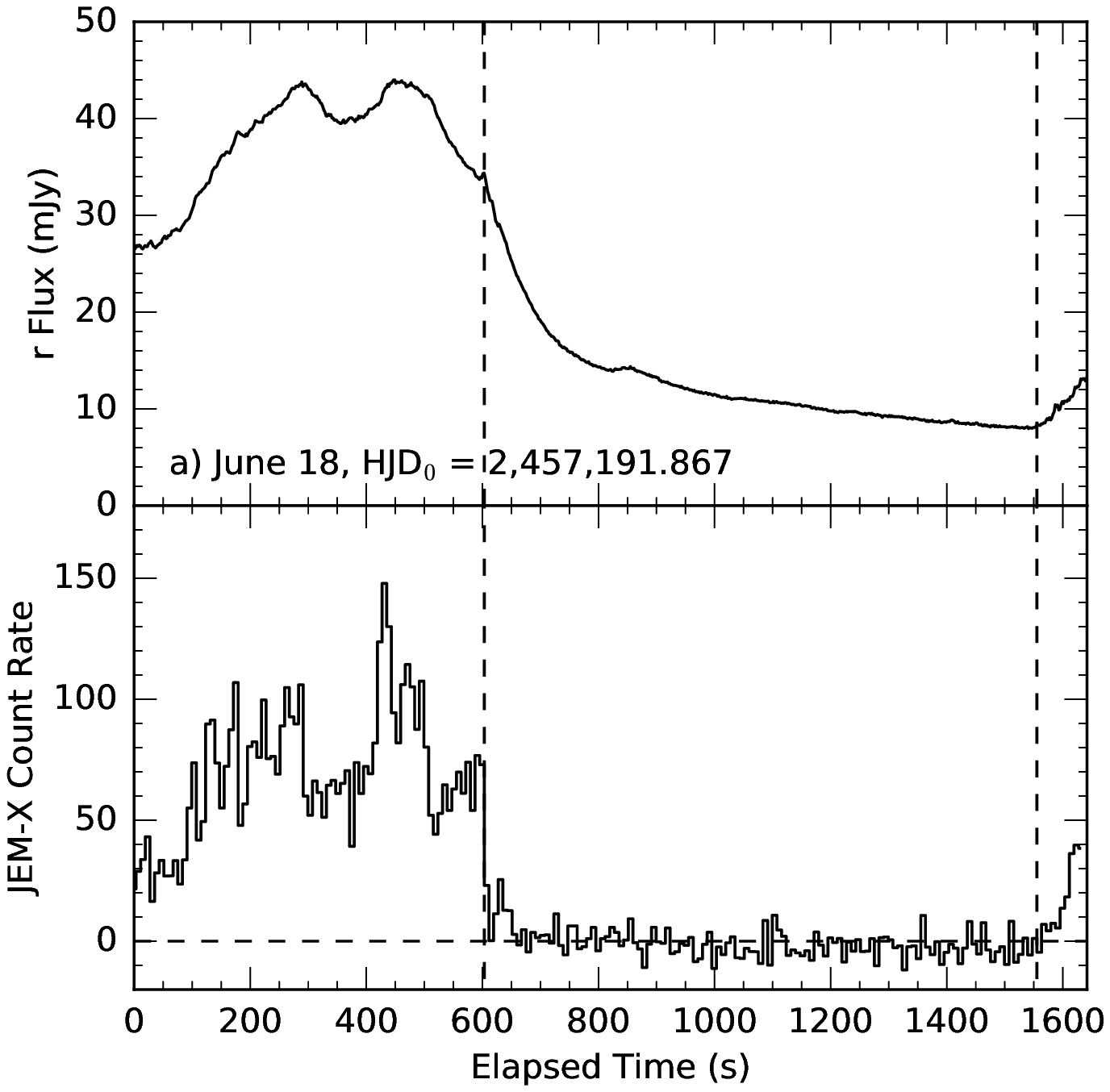}
\includegraphics[width=3.4in]{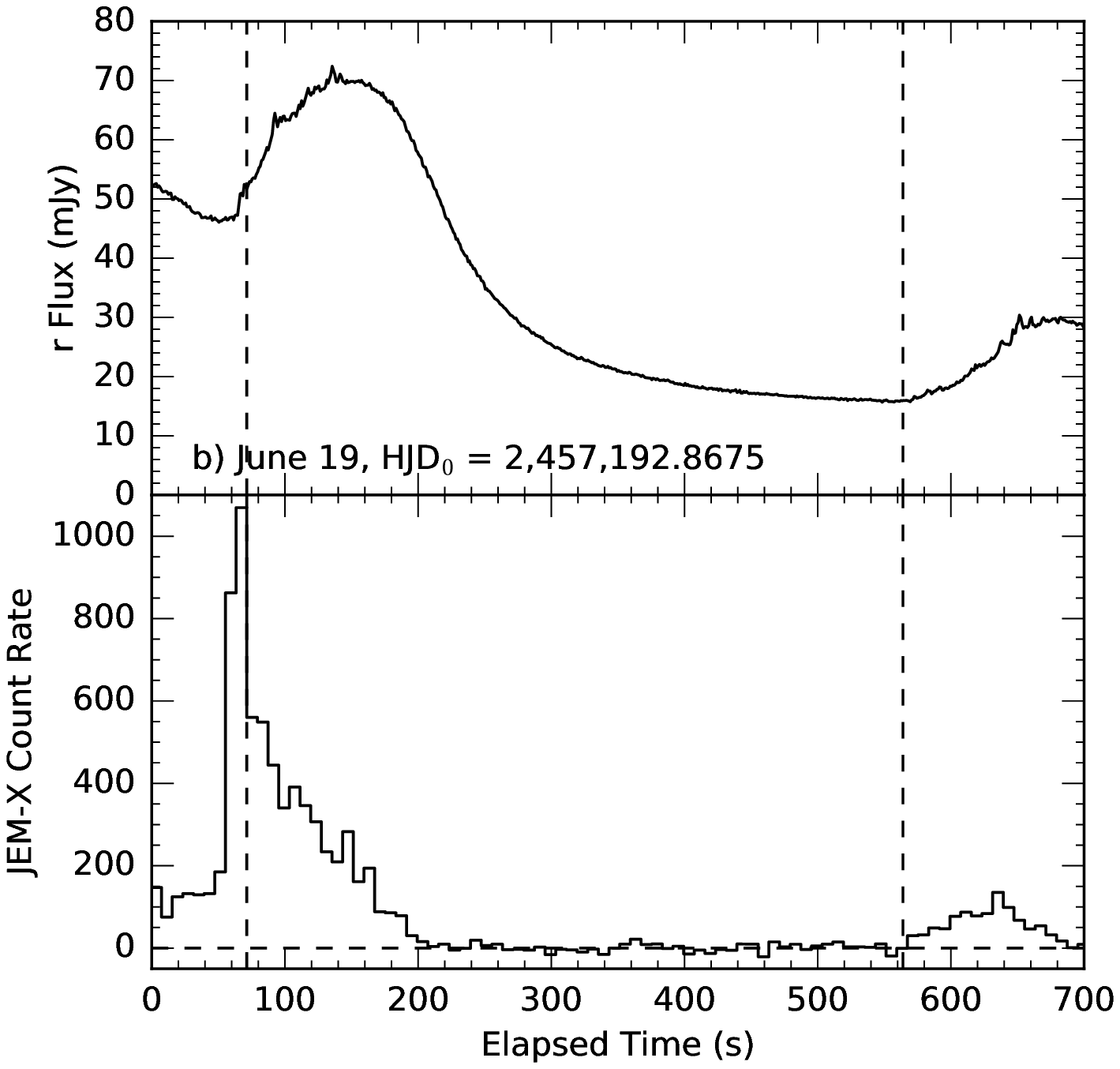}
\caption{The two most pronounced examples of smooth optical decays
  after sharp X-ray drops.}
\label{ExpFitFig}
\end{figure*}

On a number of occasions indicated by A in Figs.~\ref{McDFig} \&
\ref{MDMFig} we see rather smooth optical decays with a
near-exponential form. These can extend over more than a magnitude in
brightness and last for $\sim5-15$ minutes. On every occasion they
correspond to dramatic drops in the X-ray and gamma-ray flux to
undetectable levels. We show expanded views of the two most pronounced
events in Fig.~\ref{ExpFitFig}. Both are taken from McDonald data, one
from June 18 and the other from June 19. On June 18, both X-ray and
optical lightcurves are complex, but on June 19, the X-rays are
dominated by a single pronounced peak. The optical then peaks 1--2
minutes later.  On June 18 we see the decay sustained for 1000~s, and
on June 19 for 400~s, both much longer than light travel times could
account for. Similar behaviour is seen in multiwavelength observations
by \citet{Tetarenko:2017a} and examined in detail by
\citet{Dallilar:2017a}; we will discuss this further in
Sections~\ref{JetSection} and \ref{WindSection}.

\subsection{Optical Flaring Transitions}
\label{OpticalTransitionSection}

\begin{figure}
\includegraphics[width=3.4in]{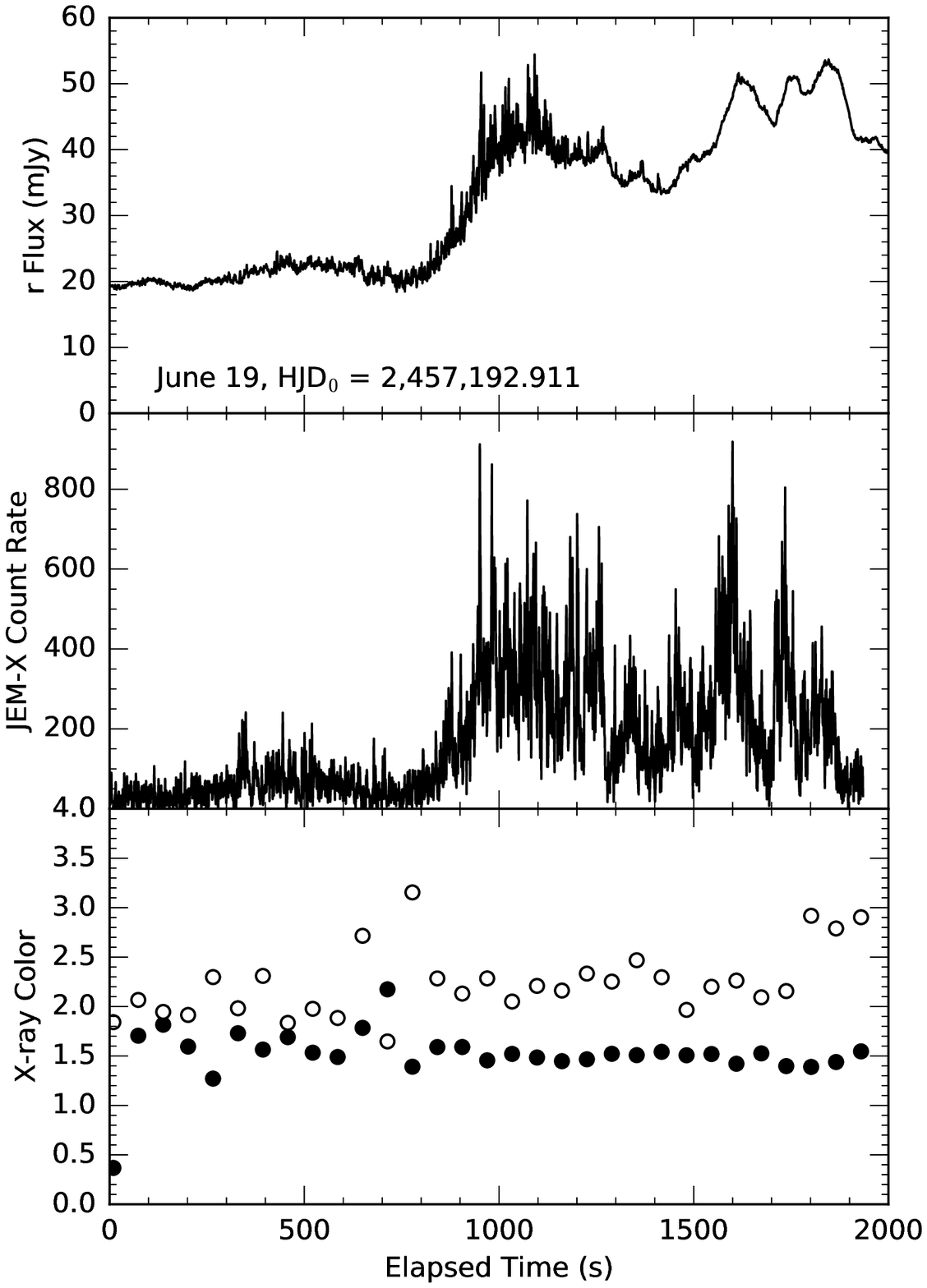}
\caption{Example of a quick transition in the optical flaring
  behaviour. seen by McDonald on June 19. The JEM-X 5--25~keV
  lightcurve at 1~s resolution is plotted for comparison,
  together with the 64~s binned evolution in X-ray colours as
  defined in Fig.~\ref{XColFig}b). White
  circles are the soft colour ([10--25~keV] / [5--10~keV]), black
  circles are the hard colour ([60--200~keV] / [25--60~keV]).}
\label{FlareTransitionFig}
\end{figure}

Another recurrent characteristic is abrupt changes in the rapid
flaring in the lightcurve. Periods of rapid flaring are signified by B
in Figs.~\ref{McDFig}--\ref{UltracamFig}, and we expand one of the
most dramatic examples in Fig.~\ref{FlareTransitionFig}.

When present the flaring can have large amplitude ($\la0.5$ mag) on
timescales of a few seconds.  We identify these as periods of
repeated flaring on timescales much less than 30~s, with amplitudes
much larger than the noise in the data.  We attempted to classify them
automatically using, for example, the r.m.s. variability within short
time bins, but found that the large variety in behaviour produced many
false positives. Instead, we selected B regions by eye, using the
local r.m.s. variability as a guide.  This variability has been
examined much more extensively by \citet{Gandhi:2016a} and
\citet{Gandhi:2017a} who find comparable or larger amplitudes which
can be unresolved at a 24~ms time-resolution.  The transition between
flaring and non-flaring optical emission can emerge in the space of
about a minute coincident with no apparent change in X-ray
behaviour. In particular, in the example shown in
Fig.~\ref{FlareTransitionFig} the 1~s resolution JEM-X lightcurves
show that X-ray flaring continues virtually unchanged after the {\em
  correlated} optical flares subside. There is no appreciable change
in X-ray colours during the transition; the hard colour is very
stable, and the soft colour shows fluctuations probably associated
with variable absorption. We will discuss this further in
Section~\ref{AbsorptionSection}.

\subsection{Discrete Optical Flaring Events}

\begin{figure*}
\includegraphics[width=3.4in]{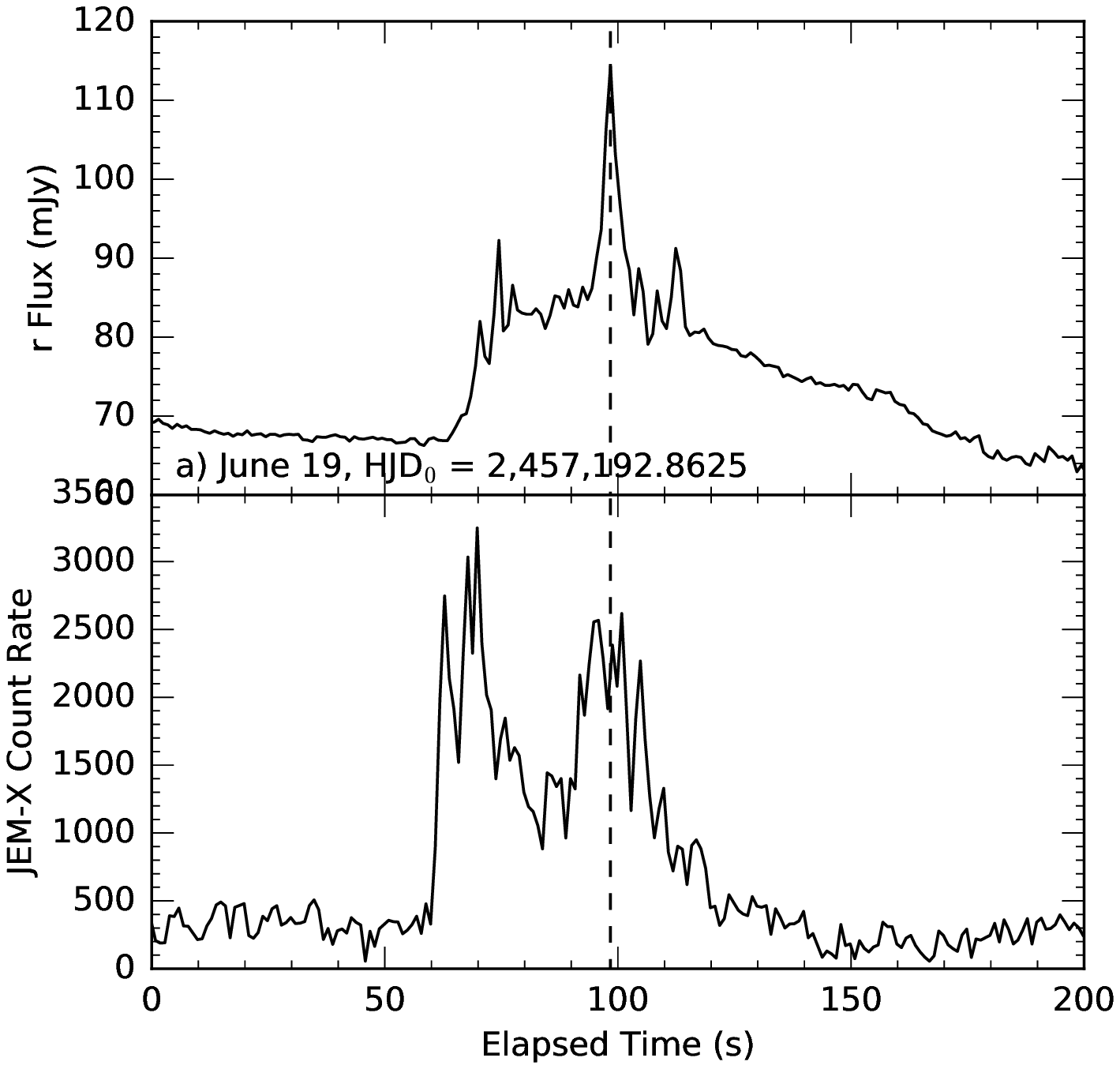}
\includegraphics[width=3.4in]{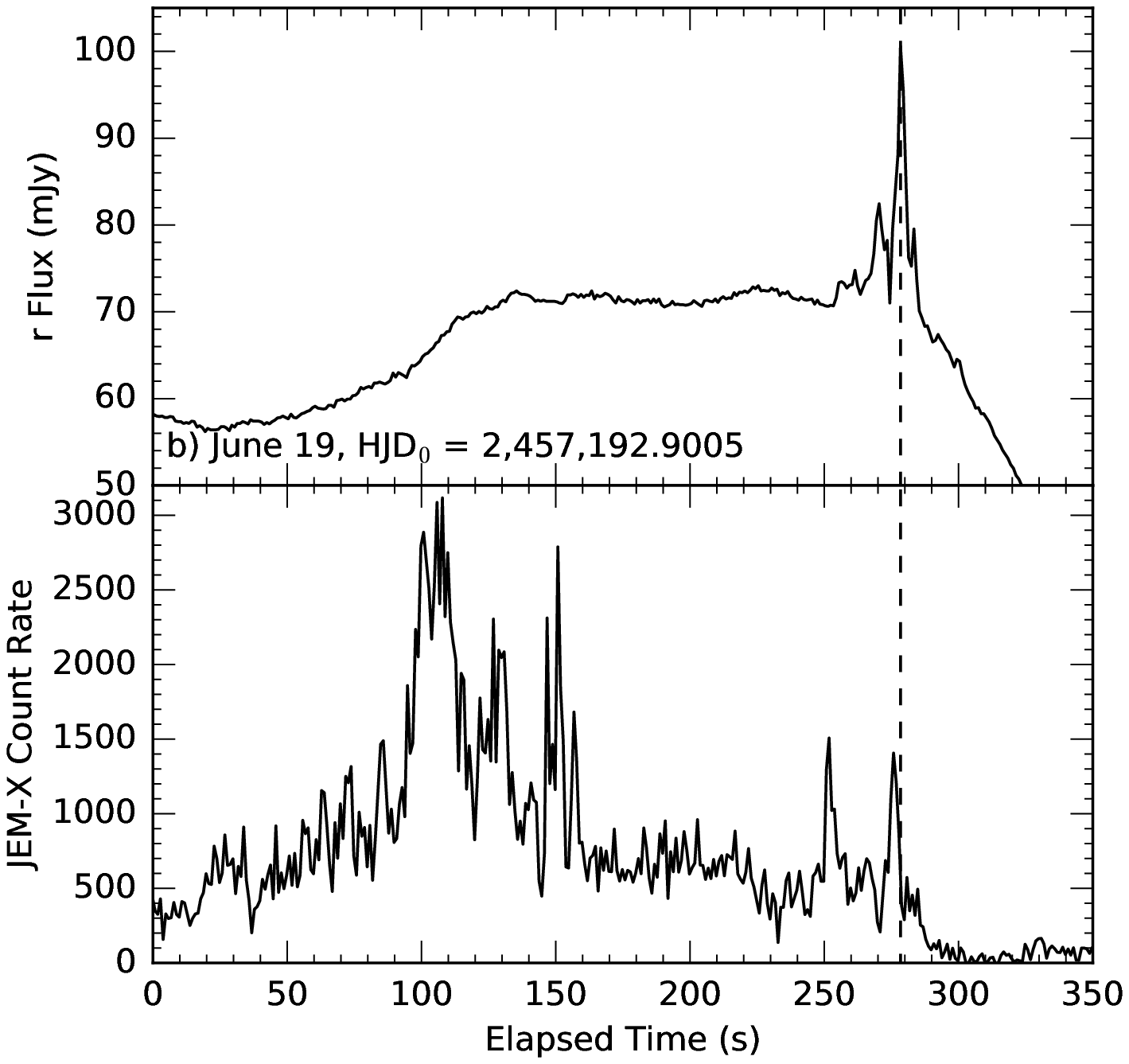}
\caption{Close-up of the two large amplitude transient optical flaring
  episodes seen by McDonald on June 19. The JEM-X 5--25~keV lightcurve
at 1~s resolution is plotted for comparison.}
\label{OptFlareFig}
\end{figure*}

Even more intriguing are several occasions, marked with C in
Figs.~\ref{McDFig}--\ref{UltracamFig}, when short-lived and discrete optical
flaring emerges. In two cases seen on June 19, expanded in
Fig.~\ref{OptFlareFig}, these occur close to the peaks of very strong
X-ray flares and are strong, peaking at $\sim40$~percent above the
persistent optical flux. These events were both seen independently by
McDonald and MDM, so are clearly real. While there is a general
association with extremely strong, sharp X-ray flares, as seen in
Fig.~\ref{McDFig}b, in detail the correspondence between X-ray and
optical behaviour is poor as shown in Fig.~\ref{OptFlareFig}. During
the first of these two events, the optical peaked about 40~s
after the strongest X-ray peak, but did coincide with a second group
of blended flares. The second optical event occurred about 170~s
after the largest X-ray peak, but may be associated with a smaller
X-ray peak a few seconds earlier. If the associations are real rather
than chance alignments, then the relative strengths of flares in
X-rays and optical are again very weakly coupled, similar to the way
in which optical flaring transitions can occur with no apparent change
in X-ray behaviour (Section~\ref{OpticalTransitionSection}).

Two other events were observed by MDM (June 20) and Ultracam (June 25)
during declines from peaks, and earlier on June 25 transient optical
flaring occurs in the middle of a relatively flat, and otherwise
unremarkable section of Ultracam lightcurve \citep[see][for discussion
of this Ultracam event]{Gandhi:2017a}. No {\it Integral} data were
available for these events, although one was observed by {\it NuSTAR}
\citep{Gandhi:2017a}.

The transient flaring properties seem similar to those during the more
extended periods described in the previous section, although the first
three events are dominated by a single extremely strong optical
flare. The proximity with strong X-ray flares in two of the events
suggests a connection, even though the largest X-ray and optical
flares are not simultaneous. While the last event observed by Ultracam
occurred in a relatively flat optical phase, X-ray coverage at the
beginning of this lightcurve indicates that X-ray flaring was
occurring, and not producing pronounced effects on the optical
lightcurve. It therefore remains quite possible that all four events
occurred close in time to sharp X-ray flares. \citet{Gandhi:2017a}
suggest that the weak Ultracam event marks a state transition between
the quiescent early part of the June 25 observation and the much more
active later phase.

\section{X-ray and Optical Flux Correlations}
\label{FluxCorrelationSection}

\begin{figure*}
\includegraphics[width=7.0in, trim=0.5in 0.0in 0.0in 0.0in, clip]{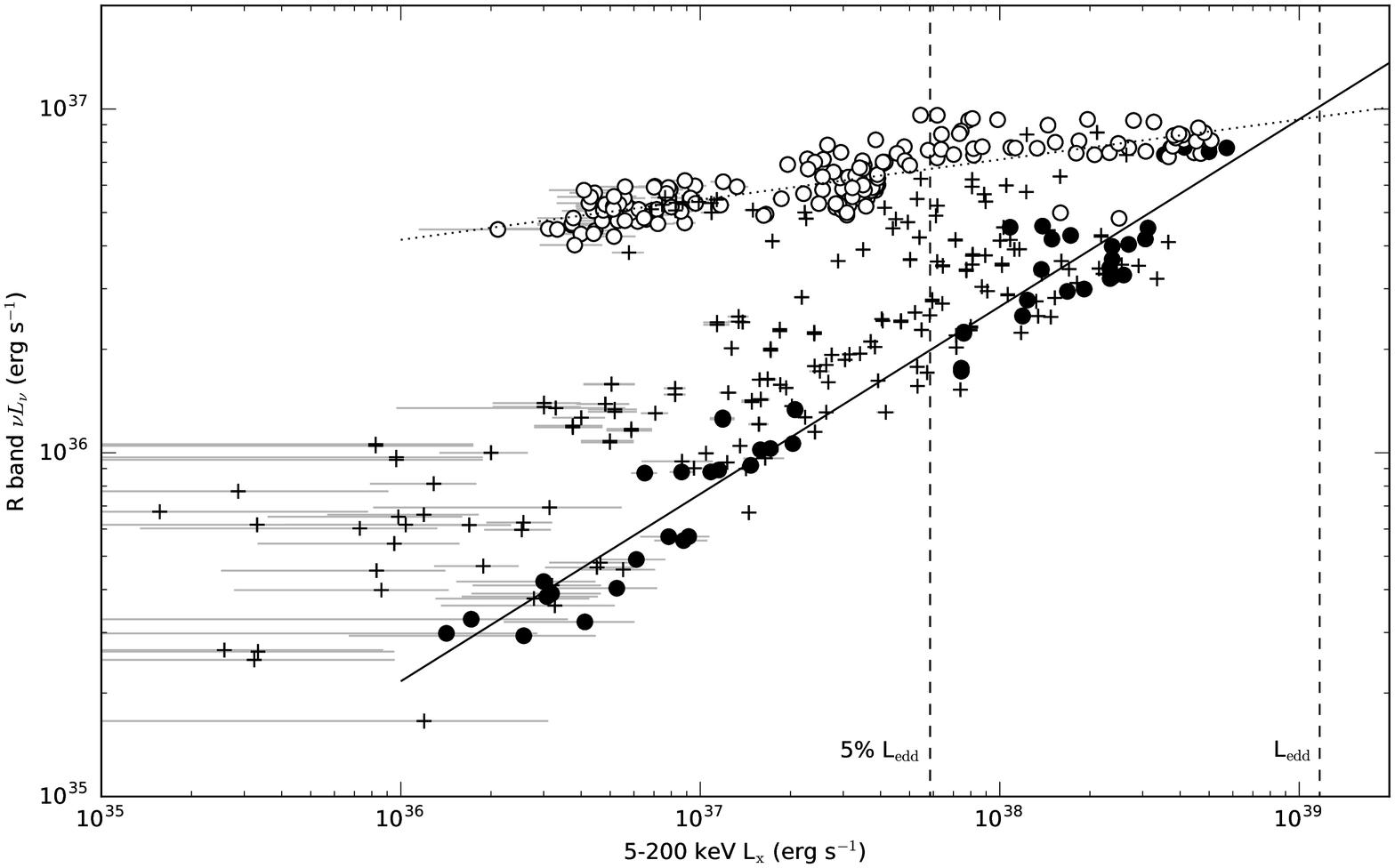}
\caption{Optical vs. X-ray/gamma-ray luminosity during outburst. Solid
  circles are times when strong rapid optical flaring was
  present. These are the periods identified by B in
  Section~\ref{OpticalTransitionSection}. The solid line, with slope
  0.55, is a power-law fit to this subset of data. Open circles are
  times when little optical variability was present and comprise all
  of the MDM data from June 21, the second half of the MDM data from
  June 22, and WHT data from June 25. The dotted line, with slope
  0.12, is a power-law fit to this subset of data. The remaining
  optical data that do not fall clearly into either of these
  categories are marked with pluses.}
\label{FluxCorrFig}
\end{figure*}

We compare inferred X-ray and optical luminosities in
Fig.~\ref{FluxCorrFig}.  Each point is based on summing average
JEM-X and ISGRI fluxes during a 200~s bin, and converting to an
(absorbed) isotropic source luminosity for a 2.39~kpc
distance. Similarly, the optical flux density during the 200~s
bin is converted to an estimate of optical luminosity in $\nu L_{\nu}$
form. X-ray to optical lags of order a minute or less inferred in
Section~\ref{EchoSection} are smaller than the time-bin used here so
have negligible effect on the plotted relationships.

\begin{figure}
\includegraphics[width=3.4in, trim=0.0in 0.0in 0.0in 0.0in, clip]{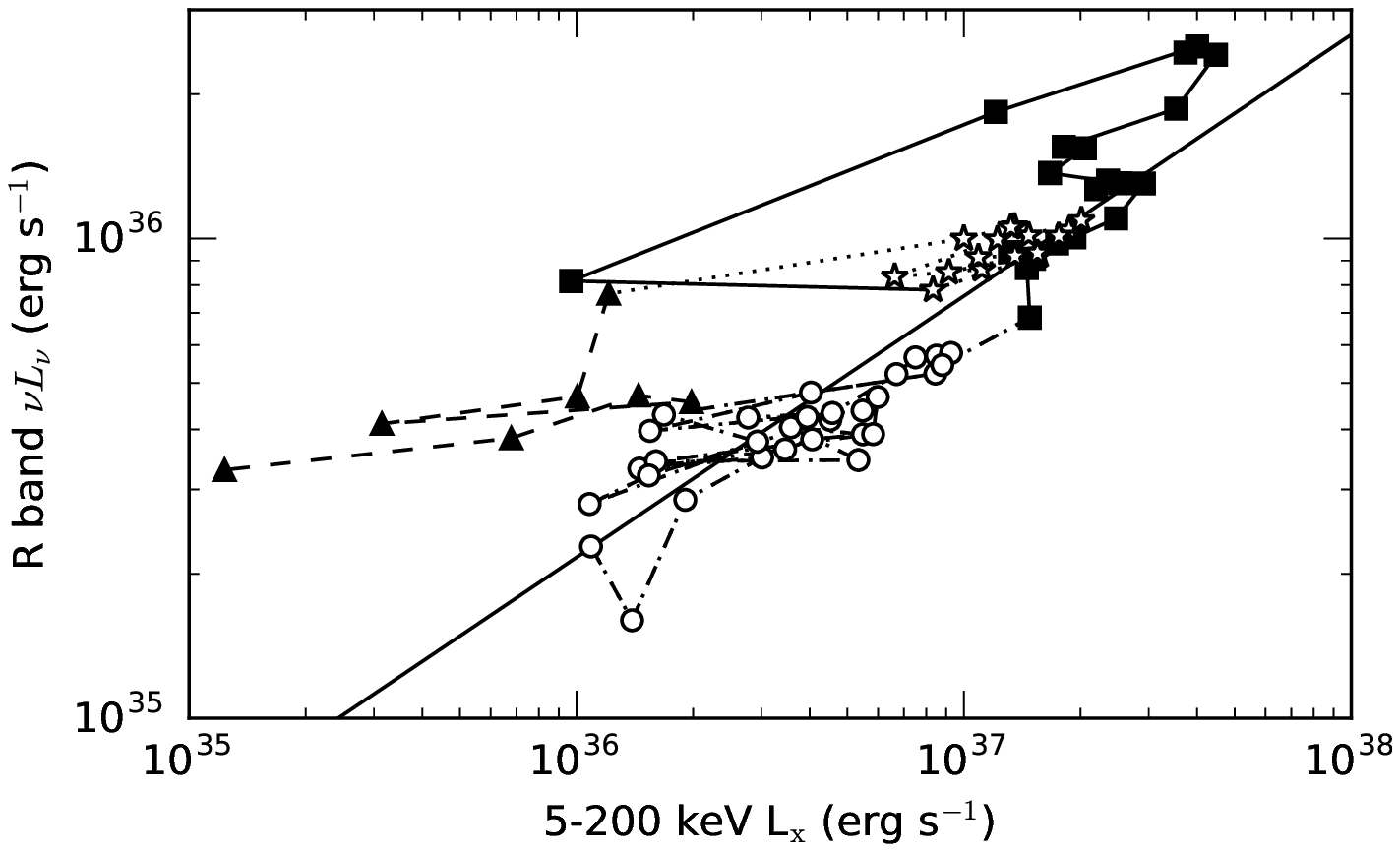}
\caption{Optical vs. X-ray/gamma-ray luminosity based only on June 18
  data from McDonald. Four sections are identified, with times
  expressed as HJD$_{\rm UTC}-2,457,190$ as in Fig.~\ref{McDFig}:
  before time 1.845 (open circle, dot-dash line), 1.845--1.88 (filled
  square, solid line), 1.88-1.91 (open star, dotted line), after
  1.91 (filled triangle, dashed line).}
\label{FluxCorrSubFig}
\end{figure}

We see correlations between X-ray and optical fluxes within each
observation. Furthermore, several broader patterns across observations
emerge. Firstly we note that the lower envelope of all of the
observations traces a quite distinct power-law. This is followed over two
orders of magnitude in X-ray luminosity. Secondly a distinct set of
observations at high optical fluxes appear to trace a flatter line,
forming an upper envelope.

We can define these subsets more clearly by examining variability in
the optical lightcurves. First we select fluxes corresponding to
periods of rapid optical flaring labelled B in
Section~\ref{OpticalTransitionSection}. These points, shown with solid
circles in Fig.~\ref{FluxCorrFig} are all clustered
along the lower envelope. We fit a power-law to these data points and
obtain a power-law index of $0.54\pm0.02$, i.e. $\nu L_{\nu} (R) \propto L_{\rm
  X}^{0.54}$. The Pearson linear correlation coefficient between the
logarithmic fluxes is 0.97. We test the significance of this
correlation using a Monte Carlo permutation test. We preserve ordering
within each night's data to account for strong auto-correlations
within the data, but  randomly reorder the nights amongst
themselves. For reach trial resampling we independently reorder X-ray
and optical data in this way, and then also apply a random cyclic
shift between the two datasets, and recalculate the correlation
coefficient. We finally determine the fraction of these permutations
which produce a correlation (or anti-correlation) of magnitude larger
than that observed. In this case the fraction of trials with an
absolution correlation larger than 0.97 (i.e. the probability of such
a strong correlation arising by chance) is $2\times10^{-4}$, so this
relationship appears to be robust.

The clumping around the upper envelope is harder to define
within small subsets of data, but seems to be epitomized by MDM data
from June 21 (when a large X-ray flare produces no pronounced optical
response), from the second half of the MDM data on June 22, and
Ultracam data from June 25. We selected these lightcurves by
  choosing those which show the weakest optical variability, sometimes
  even when large amplitude X-ray variations occur, as was notably the
  case on June 21. We highlight data from these three nights with
open circles in Fig.~\ref{FluxCorrFig}. A power-law fit to these data
points yields a power-law index of $0.12\pm0.01$ and a Pearson
correlation coefficient of 0.80, with about a 4~percent probability of
a correlation this strong arising by chance. There definitely
  appears to be a clumping in the diagram around the upper
  envelope. The value of the correlation coefficient does, however,
  leave the possibility that this is not truly tracing a rising trend
  with increasing X-ray luminosity. Instead it is possible that
  different observations with different optical brightnesses may
  happen to align to produce this trend by chance. Nonetheless, it is
  clear that the periods of weakest optical variability occur near the
  upper envelope, when the optical to X-ray flux ratio is highest.
Finally we note that the only Ultracam observation to show rapid
flaring (June 26) is also the only one to lie on the lower envelope
rather than the upper one.

The lower envelope we identify is strikingly similar to the
$L_{\rm opt} \propto L_{\rm X}^{0.61}$ correlation identified from
outburst to quiescence in a number of BHXRTs \citep{Russell:2006a},
and that specific to V404~Cyg collated by \citet{Bernardini:2016a}:
$L_{\rm opt} \propto L_{\rm X}^{0.56}$. As done by
\citet{Bernardini:2016a}, we can extrapolate the lower envelope fit to
compare to quiescent data from \citet{Hynes:2009a}, after subtracting
the orbital-phase modulated companion star flux to leave the residual
accretion emission. We do not show this in Fig.~\ref{FluxCorrFig} to
avoid compressing the outburst data, but we do find the extrapolation
falls within about a factor of two of the quiescent accretion light.

\citet{Kimura:2017a} examined X-ray vs. optical correlations during
the 2015 December outburst in a similar way, and found power-law
indices of 0.25--0.29. These are based on much more limited time
coverage than we have used, however, and sample a smaller dynamic
range of luminosity. They considered a single relation,
rather than looking at envelopes of a more complex behaviour, and
found a power-law index intermediate between our two cases. Their
results then appear broadly consistent with the range of behaviour we
observe. 

As the inferred X-ray luminosity increases the two envelopes
converge. With a little extrapolation, the power-law fits actually
join very close to the Eddington limit ($\sim0.85L_{\rm Edd}$). Given
uncertainties in distance and black hole mass, and the intrinsic
geometric uncertainties inherent to applying the Eddington limit to a
disc accreting system, this is quite a good agreement.  This apparent
convergence is borne out by the group of points seen at the highest
inferred luminosity where we see both optically flaring (i.e. lower
envelope), and low variability (i.e. upper envelope) segments of data
coincident in the diagram.

The correlation between X-rays and optical {\em along} the envelope is
most pronounced when the uncorrelated optical component is minimized
on June 18. We expand the correlation for this night in
Fig.~\ref{FluxCorrSubFig}, dividing it loosely into four phases, for
which we refer to times $t={\rm HJD}_{\rm UTC}-2,457,190$
corresponding to the time axis in Fig.~\ref{McDFig}. During the first
phase, $t<1.845$, the source lies just above the envelope, and moves
up and down roughly parallel to it. The X-rays are continuously
present, and both they and the optical emission maintain low level
variability. During the major flare of this observation,
$1.845<t<1.88$, the fluxes move parallel to the envelope to the
highest fluxes reached on this night, before entering an X-ray dip
where the X-ray flux dramatically decreases, with optical flux
decaying smoothly, and never quite even reaching the level seen at the
beginning of the night. This has the effect of moving the source away
from the envelope, with the optical appearing independent of the
instantaneous X-ray brightness. After the dip, X-rays resume activity
during $1.88<t<1.91$, moving the source back to the envelope,
overlaying the position occupied earlier in the night. Finally for
$t>1.91$, the X-rays enter a more extended dip, allowing the optical
to more fully decay back to levels seen at the beginning of the
night. At this point, the X-rays are much fainter than initially seen
so the source has again moved well away from the envelope. The optical
then appears to be sustained at a level unconnected to the X-ray
flux. Detailed examination of this observation, then, shows that the
envelope is followed within a night, and not just from night to night.

Finally, we note that the two states seen in the X-ray vs. optical
relations are quite different to the behaviour of most BHXRTs
described by \citet{Russell:2006a} which are optically fainter in
softer states. This behaviour was, however, seen in the 1989 outburst
of V404~Cyg, manifesting as high outlier points, and seems to be a
characteristic unique to, but repeatable in, this source.

\section{Associations Between Optical Variability and X-ray Colours}
\label{XrayColourSection}

\begin{figure}
\includegraphics[width=3.2in]{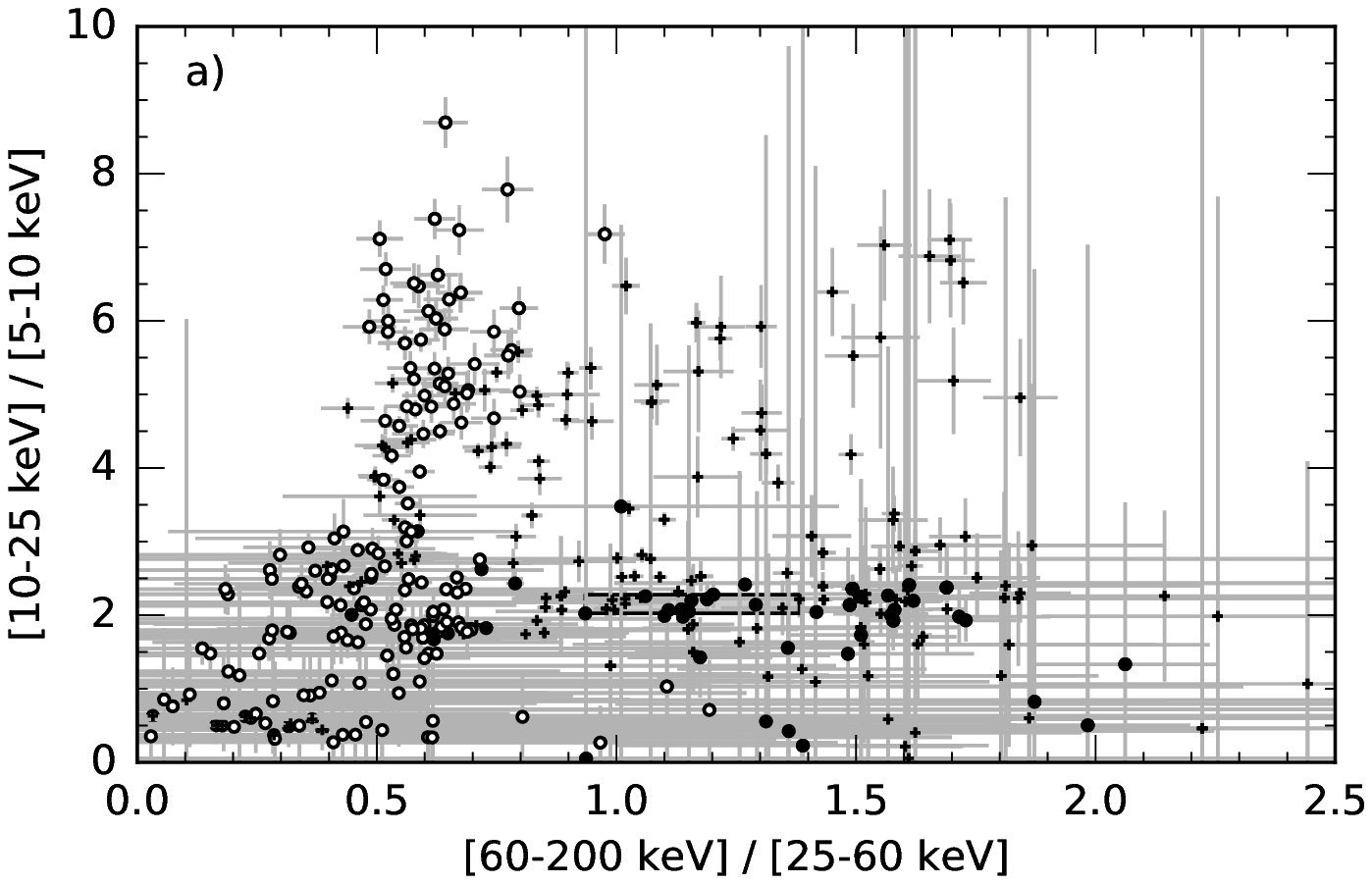}
\includegraphics[width=3.2in]{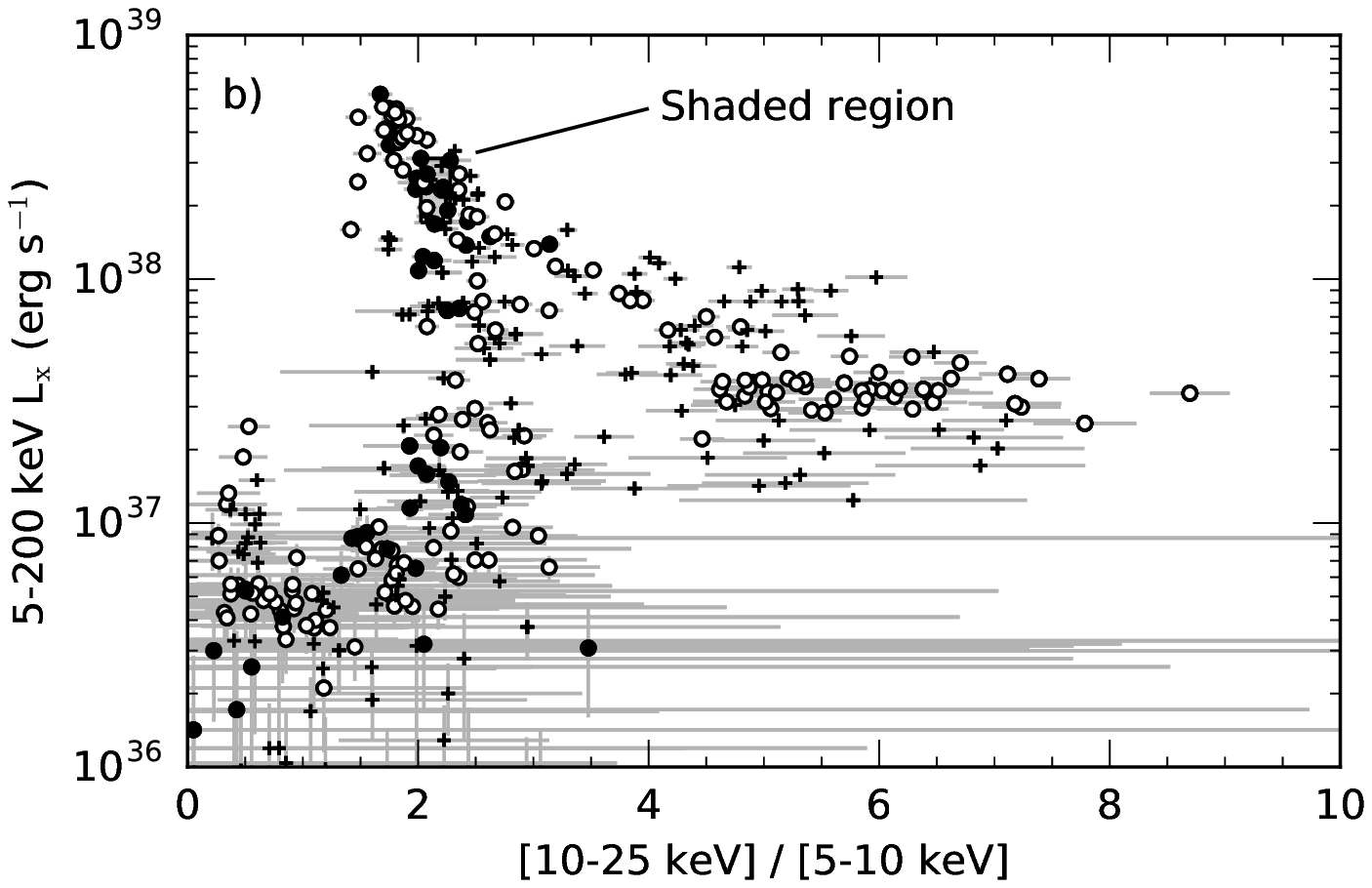}
\includegraphics[width=3.2in]{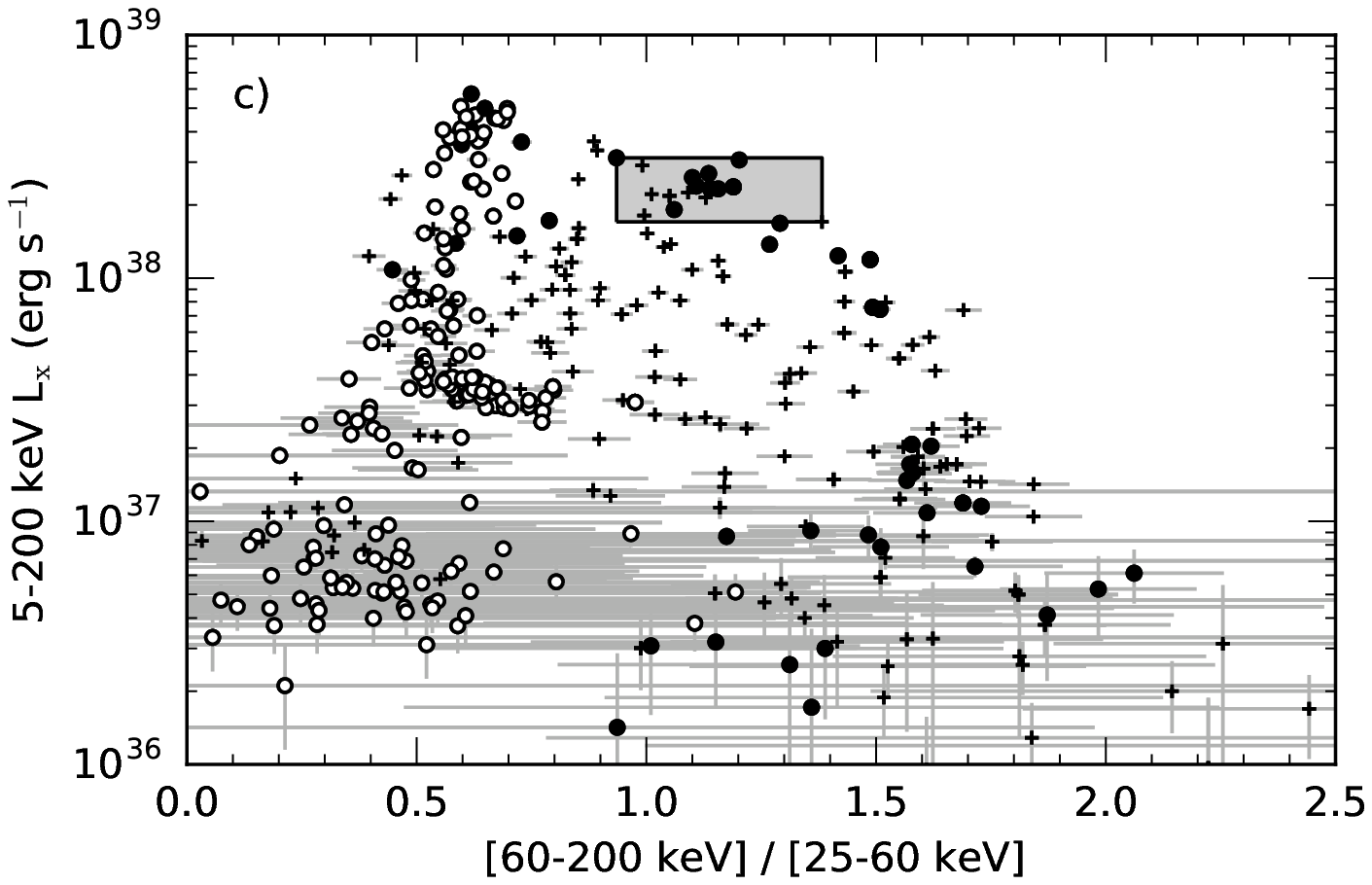}
\caption{{\em Integral} colour-colour and hardness-intensity diagrams
  at the times overlapping with optical observations. Symbols
  correspond to those used in Fig.~\ref{FluxCorrFig}, with filled
  circles indicating bins showing rapid flaring when the source was on
  the lower-envelope of the X-ray/optical diagram, and open circles
  indicating the optically bright upper envelope when little optical
  variability was present. The shaded grey area corresponds to the
  time when Ultracam saw persistent, intense, rapid flaring
  \citep{Gandhi:2016a}. In the panel (b) (the soft hardness-intensity
  diagram), the shaded area is obscured by a group of points at (2.15,
  $2.3\times10^{38}$).}
\label{XColFig}
\end{figure}

We have identified two distinct subsets of optical--X-ray/gamma-ray
behaviour which define boundaries in the flux relationships. The
lower-envelope is roughly traced by the
$L_{\rm opt} \propto L_{\rm X}^{0.56}$ line and sometimes shows rapid
($<30$~s) optical flaring. The optically bright upper envelope
approximately follows $L_{\rm opt} \propto L_{\rm X}^{0.12}$ with the
optical weakly variably and almost uncorrelated with X-rays. We can
now examine how these relate to X-ray color variations. We show in
Fig.~\ref{XColFig} colour-colour and hardness-intensity diagrams based
on {\it Integral} data at the time of optical observations.  Note that
the bands used (see Section~\ref{IntegralSection}) are skewed to
higher energies than usual colour-colour and hardness-intensity
diagrams \citep[e.g.][]{Dunn:2010a} due to the sensitivity range of
{\it Integral}, and have been chosen to highlight the difference
between the two categories of optical observations. In particular, we
see the most dramatic differences between upper and lower envelope
data when considering a hard X-ray (ISGRI) colour.

\begin{figure}
\includegraphics[width=3.2in, trim=0.0in 0.0in 0.0in 0.0in, clip]{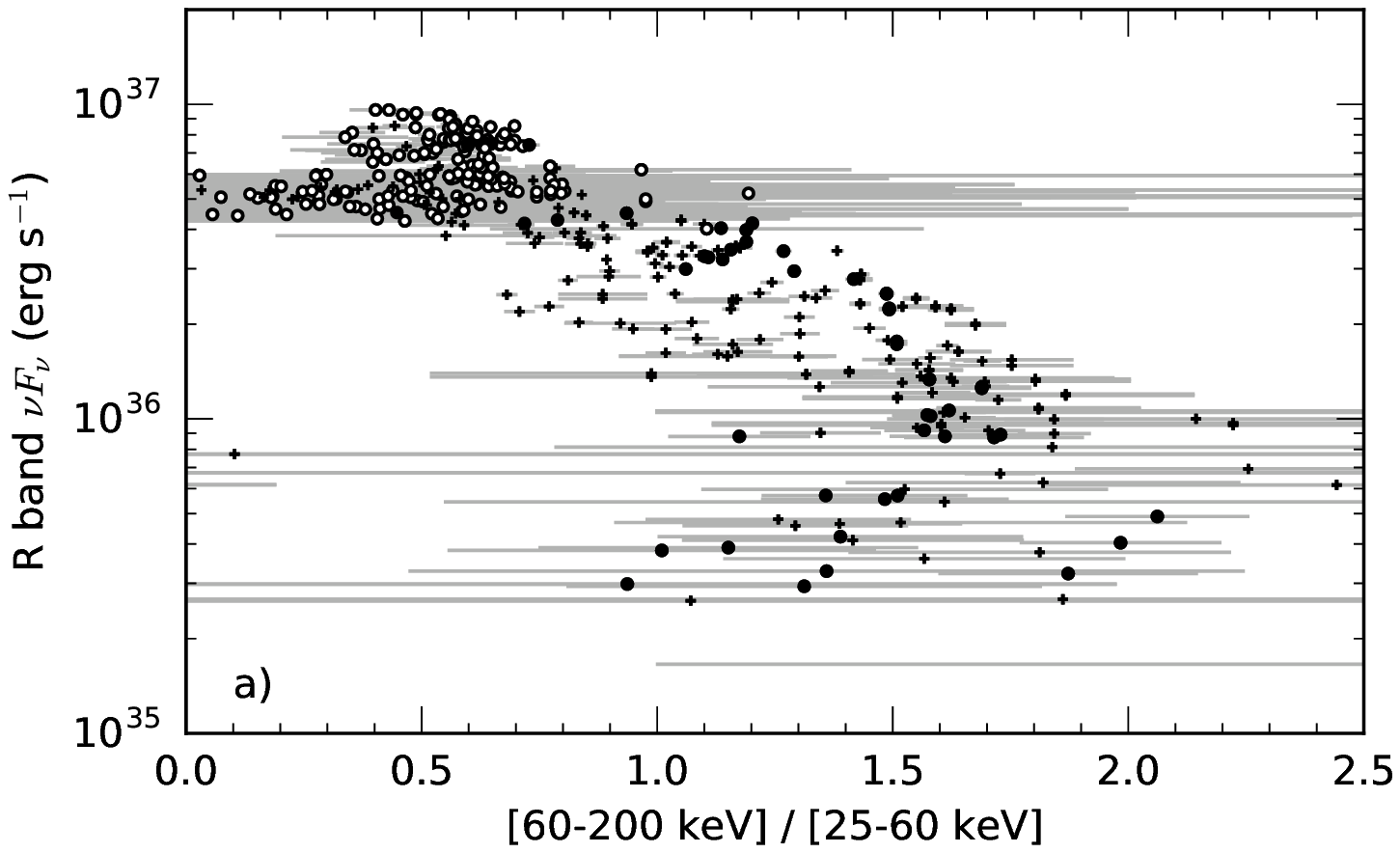}
\includegraphics[width=3.2in, trim=0.0in 0.0in 0.0in 0.0in, clip]{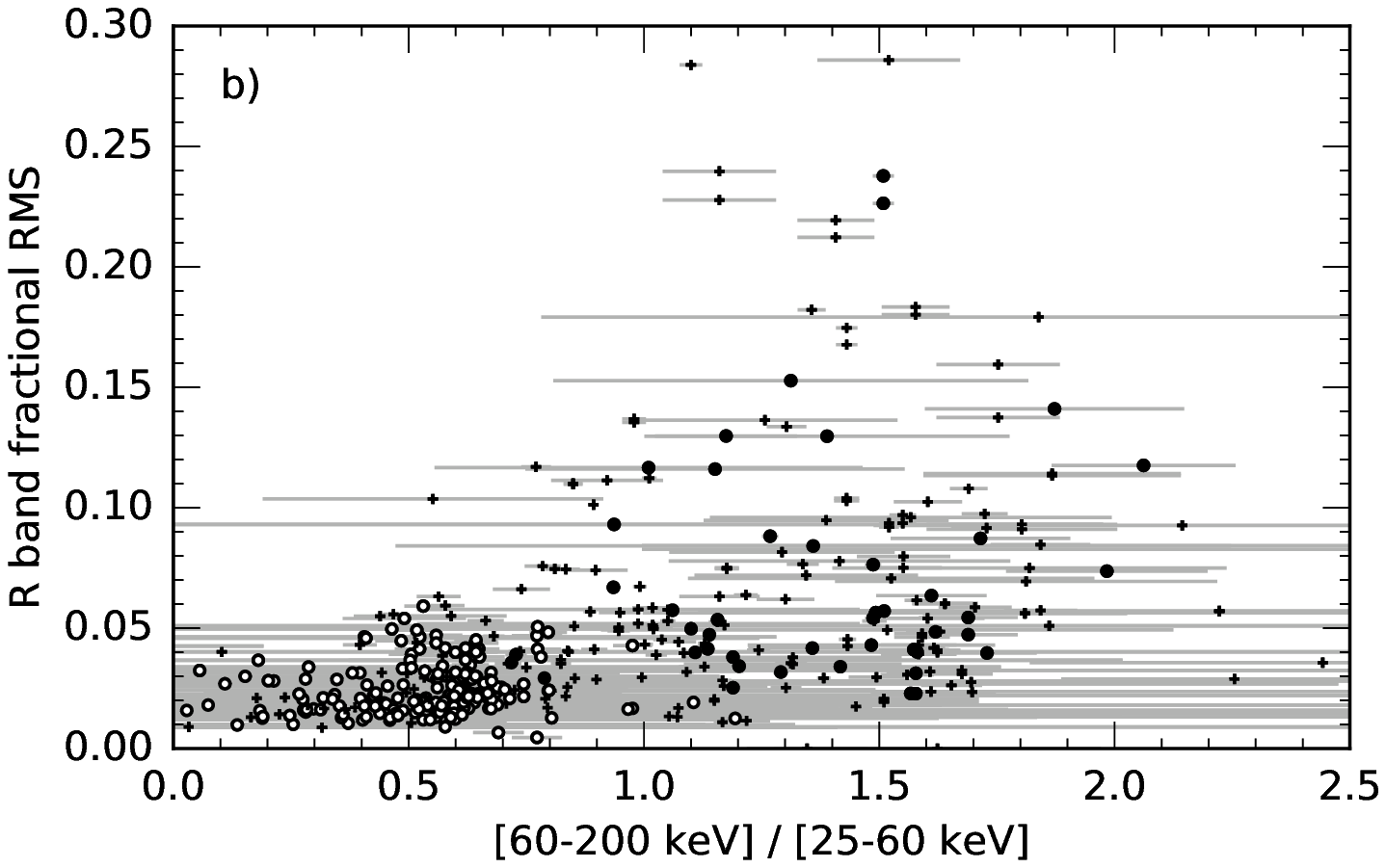}
\caption{Optical luminosity (upper), and fractional r.m.s. within a
  200~s bin (lower) as function of hard X-ray hardness. These
  quantities are clearly related, with the relation between hard
  colour and optical luminosity much clearer than that with soft
  colour. Generally, states with harder X-ray spectral shapes are
  associated with fainter and more variable optical
  lightcurves. Symbology is the same as in Fig.~\ref{FluxCorrFig}}
\label{OptColFig}
\end{figure}

We have differentiated a subset of points corresponding to the lower
and upper envelope points in the X-ray vs. optical luminosity plot
(Fig.~\ref{FluxCorrFig}). The differentiation is seen most clearly
in the hardness-intensity diagram using the hard (ISGRI) colour, where
we see that the lower-envelope is characterized by a much harder hard
colour than the upper envelope, indicating an excess of flux above
60~keV (or a deficit in the 25--60~keV range). This holds across a
wide spread of X-ray luminosities, although the two sets of points
converge in hard colour as the Eddington limit is approached (as the
two envelopes also converge in Fig.~\ref{FluxCorrFig}). We test this
differentiation by performing a Kolmogorov-Smirnoff test to check if
the distribution of hard colours seen in the lower and upper branches
could be consistent with the same distribution. As expected from the
colour-colour diagram, this is strongly ruled out with (formally) a
$10^{-21}$ probability that they are drawn from the same distribution
of X-ray colours.

The softer hardness-intensity diagram (using the JEM-X colour) shows
two tracks, diverging from the highest luminosity points. The upper
one decreases in apparent luminosity while becoming substantially
harder, and probably represents variability induced by changing
absorption. No periods of rapid optical flaring appear to be
associated with this track.  The lower track decreases in apparent
luminosities while remaining quite soft in the JEM-X band. These most
likely involve real decreases in underlying X-ray luminosity, rather
than just changes in line-of-sight absorption. The second track
involves both points identified with the upper and lower
envelope. These two tracks, then, do not seem associated strictly with
the lower and upper envelopes, although the hard upper track does
appear to preclude rapid flaring. 

Finally, the colour-colour diagram contains elements of both of the
other diagrams. The lower-envelope points mainly involve changes in
the hard (ISGRI) colour, and upper envelope points mainly track changes
in the softer (JEM-X) colour.

We further examine the relationship between hard X-ray colour and
optical variability (defined by the r.m.s. variation within a 200~s
time bin) more quantitatively in Fig.~\ref{OptColFig}. This
shows that times when the hard X-ray spectrum is hardest are
associated with lower and more variable optical fluxes. These two
relationships hold across the dataset as a whole, and are independent
of any classification into lower or upper envelopes.  Many of the
phases when the ISGRI spectrum is quite hard {\em do not} involve
rapid optical flaring, so this indicates that these phases must also
be associated with the slower timescale optical variations being
fainter and more variable as well. The rapid flaring appears to be a
symptom of things going on, but is not (alone) responsible for the
increased r.m.s. when the hard colour is hardest.

Specifically, we find that the optical fluxes appear to be
anti-correlated with ISGRI colour. This does not appear to be a linear
correlation, even with respect to logarithmic fluxes, so we test its significance
using Spearman's rank-order correlation. We find a correlation
coefficient of -0.75, with a probability of producing a chance
correlation this large of about $3\times10^{-4}$. The optical
r.m.s. variability is positively correlated with hard hardness,
although the dispersion of the correlation also increases as the ISGRI
spectrum gets harder. What does seem to be true is that large values
of optical r.m.s. variability are {\em only} seen when the hard X-ray
spectrum is quite hard.  Quantitatively, we find a Spearman rank
correlation coefficient of 0.61 between the X-ray colours and optical
r.m.s. values, with a probability of chance correlation below
$10^{-5}$.

\section{X-ray/Optical Lags and Reprocessing}
\label{EchoSection}

Optical variations generally appear to be lagged and smeared with
respect to the X-ray ones. The smearing is ubiquitous, with some
exceptions for periods when rapid optical variability emerges,
e.g. June 19, and June 26. Lags are sometimes harder to see directly
in the lightcurves, as the correspondence in strength of X-ray and
optical flares is poor. A lag is most clearly visible on June 26, when
in addition to rapid optical flaring, slow variations correlate with
X-rays, but are lagged by about a minute.

\begin{figure}
\includegraphics[width=3.4in, trim=0.0in 0.0in 0.0in 0.0in, clip]{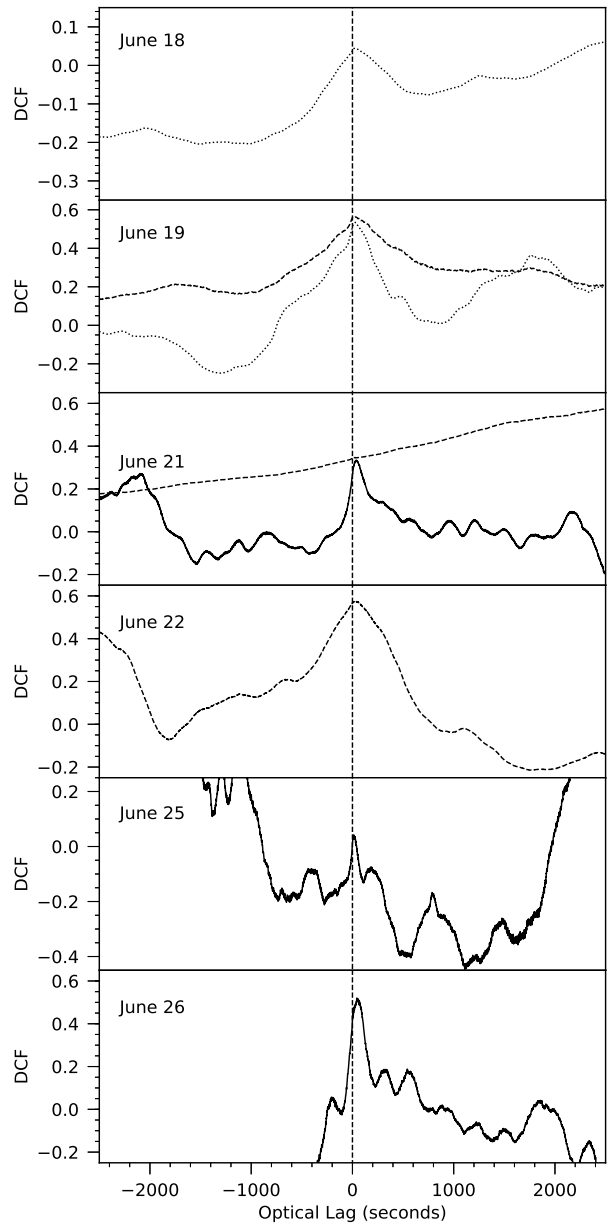}
\caption{Discrete CCFs. McDonald data are shown dotted, MDM data
  dashed and WHT data in solid lines.}
\label{AllCCFFig}
\end{figure}

\begin{figure}
\includegraphics[width=3.4in, trim=0.0in 0.0in 0.0in 0.0in, clip]{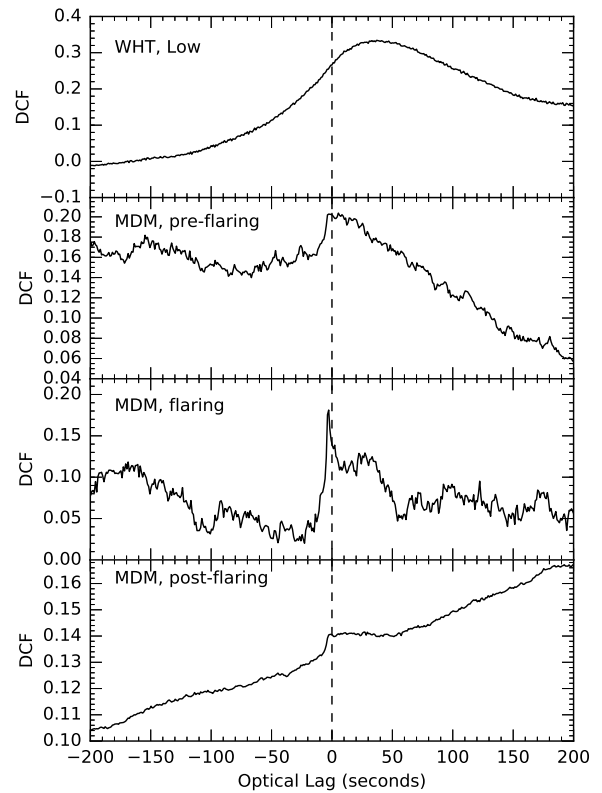}
\caption{Discrete CCFs through the large flare seen in X-rays on June
  21. The top panel uses the whole WHT dataset from that night. The
  MDM panels correspond to times 4.754--4.795, 4.795--4.820, and
  4.820--4.894 respectively.}
\label{BigFlareCCFFig}
\end{figure}

We quantify X-ray to optical correlations and their lags using
cross-correlation functions (CCFs). For this purpose we combine
optical lightcurves with JEM-X ones generated at a matching
time-resolution. We choose to construct discrete CCFs
\citep[DCFs;][]{Gaskell:1987a}, since overlaps sometimes involve
discontinuous segments. We show all available CCFs over a wide-range
of lags in Fig.~\ref{AllCCFFig}. In every case except the MDM
lightcurve from June 21 we see CCFs dominated by a single clear peak
slightly after zero lag. These are all relatively broad, although vary
substantially in width, and peak at lags of 10--50~s. This
  overlaps the 22.5~s and 34.8~s lags reported in the 2015 December
  outburst \citep{Kimura:2017a,Kimura:2018a}.  On June 21, the very
large X-ray flare which is relatively uncorrelated with optical
behaviour on long timescales dominates the MDM CCF producing a general
rise towards late times.  Close examination shows a small peak
  superposed on this rise close to zero lag (see
Fig.~\ref{BigFlareCCFFig} and discussion below).

Other studies \citep[e.g.][]{Rodriguez:2015a} have claimed that
quasi-correlated optical flaring was also seen with the optical
lagging the X-rays by anywhere from $<1$~min to 20--30~min, varying
from flare to flare. The latter observation is quite inconsistent with
our finding that the dominant X-ray/optical CCF peaks are usually
quite close to zero lag. As noted earlier, however, the lightcurves
are complex and sometimes show strong optical counterparts to weak
X-ray flares and vice-versa. This would be missed with poorly sampled
data, leading instead to matching the nearest bright flare and
inferring a larger lag than is real. Our higher time resolution
optical coverage breaks these degeneracies.

\begin{figure}
\includegraphics[width=3.2in, trim=0.0in 0.0in 0.0in 0.0in, clip]{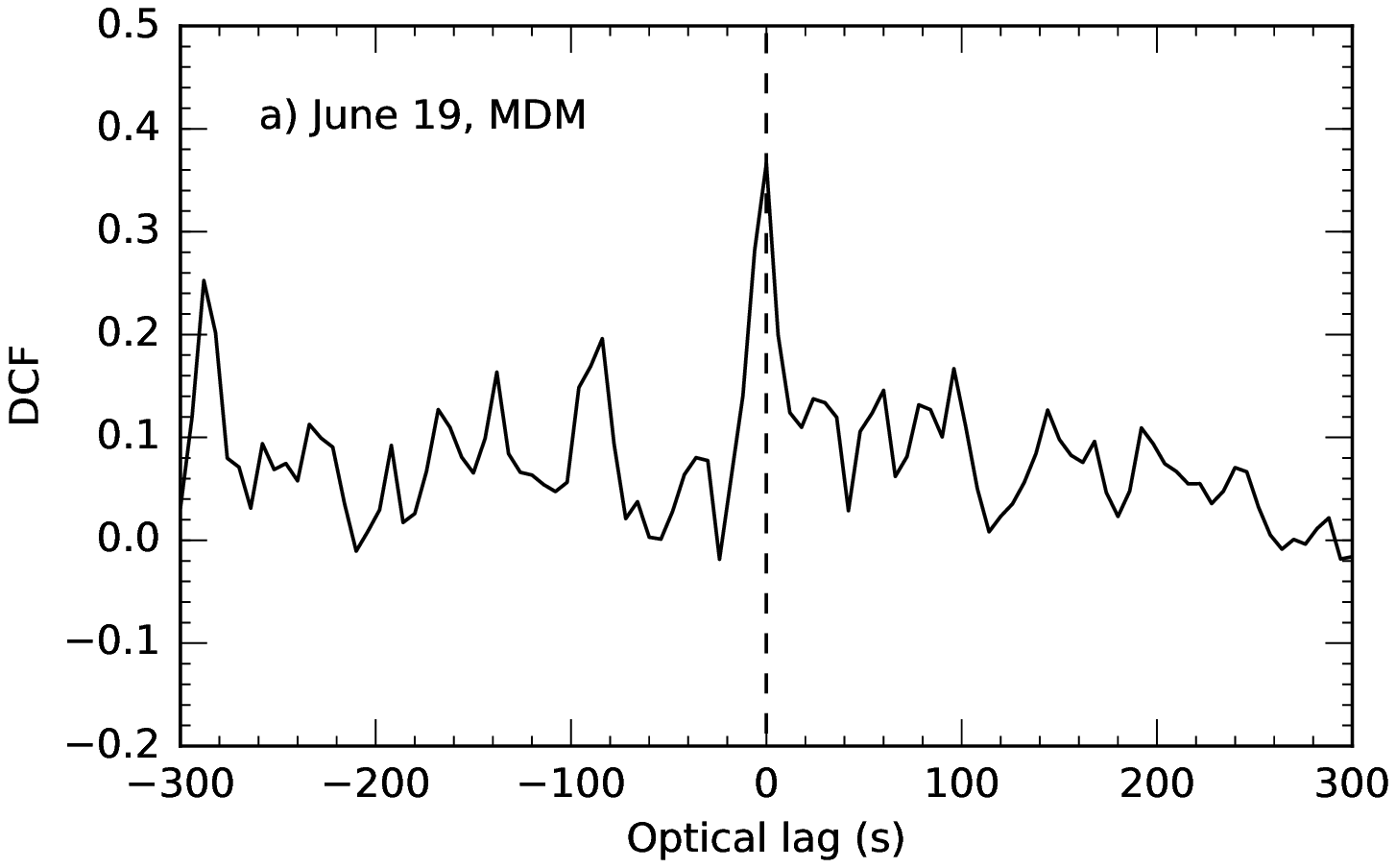}
\includegraphics[width=3.2in, trim=0.0in 0.0in 0.0in 0.0in, clip]{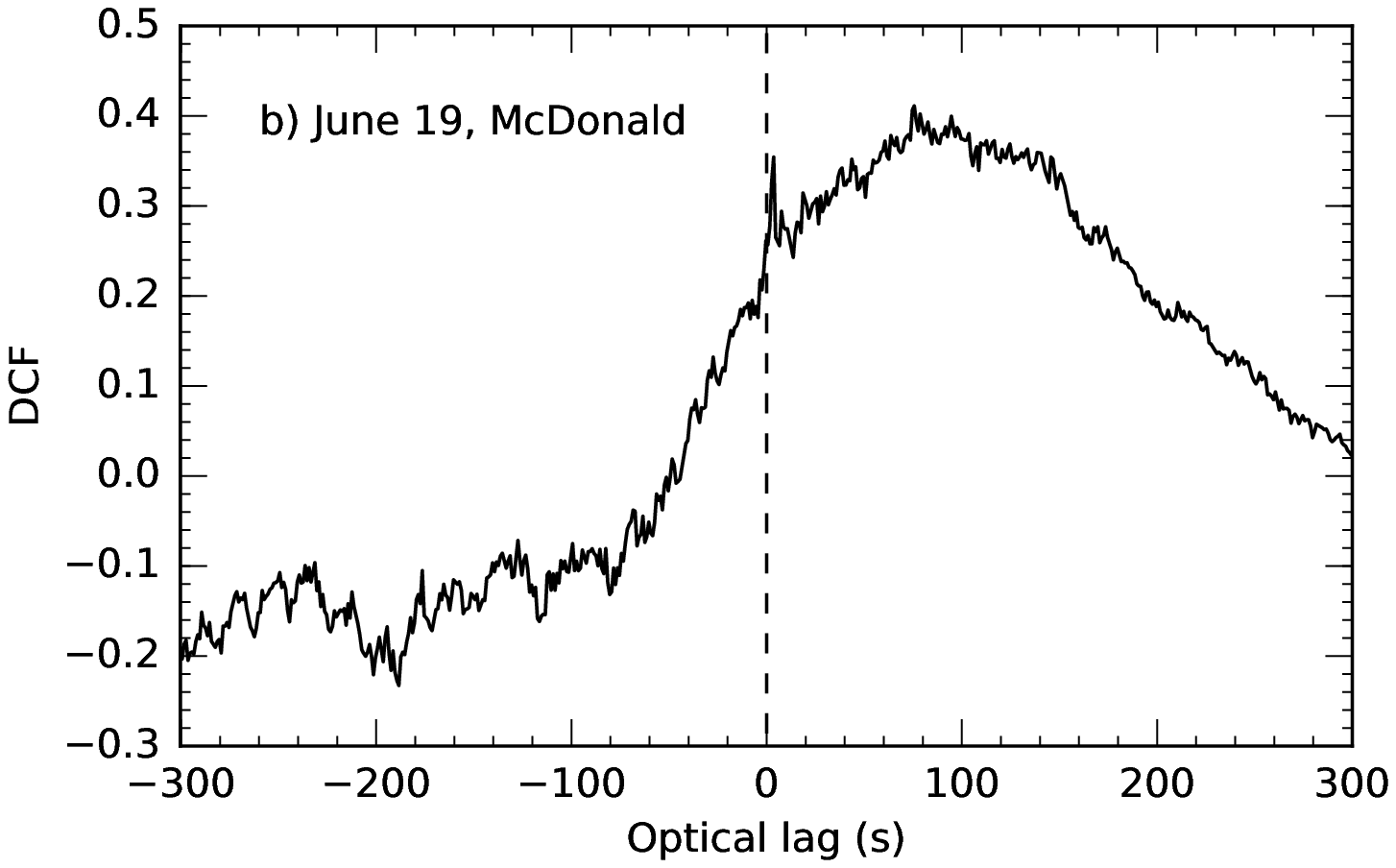}
\includegraphics[width=3.2in, trim=0.0in 0.0in 0.0in 0.0in, clip]{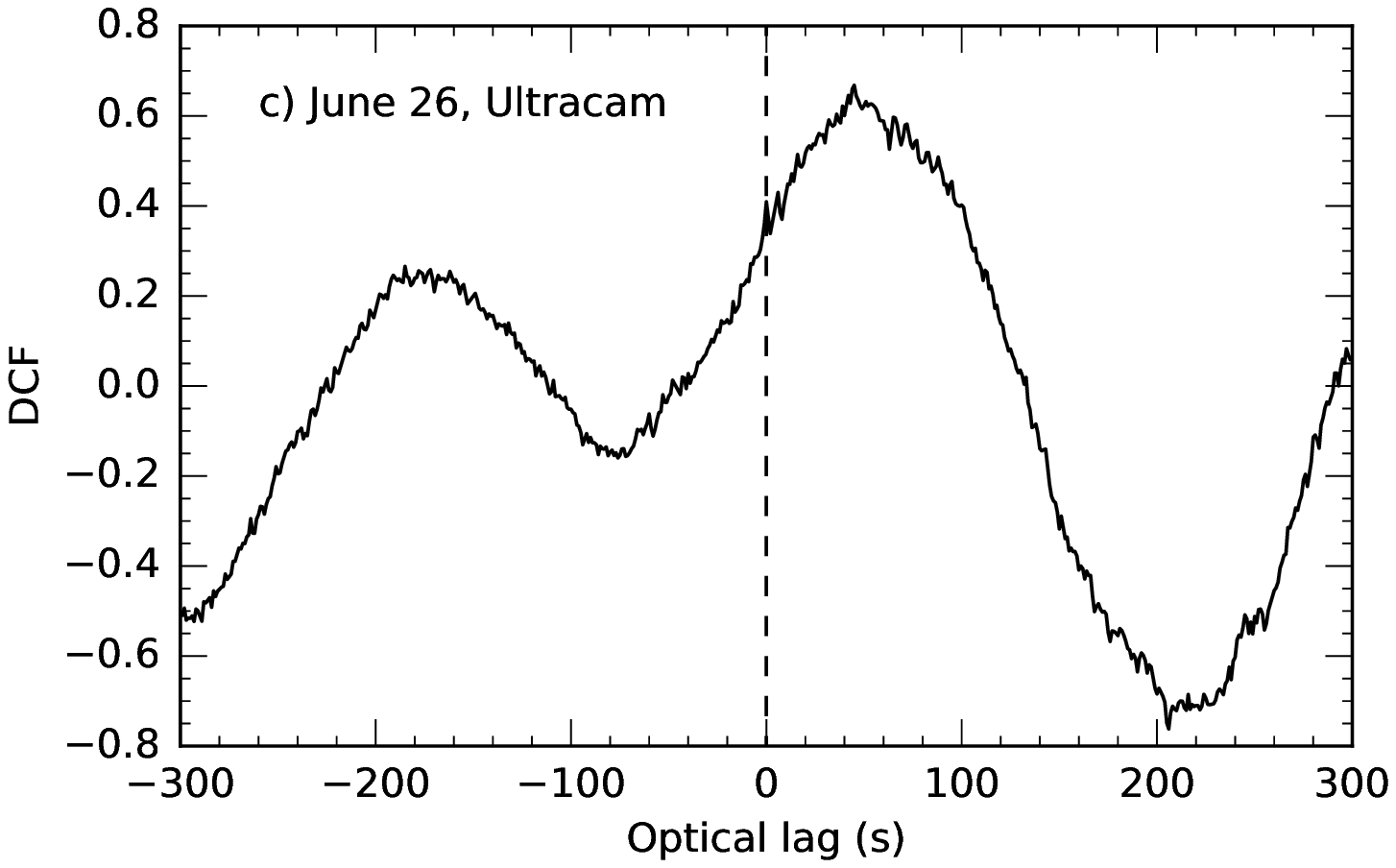}
\caption{Optical vs. X-ray cross-correlation functions during the
  rapid flaring episodes on June 19 and 26. Panel (a) is taken from
  MDM data at 6~s time resolution, towards the end of the first
  episode labelled B in Fig.~\ref{MDMFig}a, times labelled
  2.83720--2.84748. Panel (b) is taken from McDonald data at 1~s
  time resolution, during the second episode labelled B in
  Fig.~\ref{MDMFig}a (the only B episode in Fig.~\ref{McDFig}b; times
  labelled 2.921--2.926). Panel (c) uses the whole Ultracam dataset
  from June 26.  While a sharp and near instantaneous response is seen
  alone during the first episode, during the second we clearly also
  see a broad, lagged response, which appears quite distinct from the
  peak at zero lag. The June 26 data look similar to the panel (b),
  but which the sharper peak at zero lag being weaker relative to the
  large amplitude slow variations.}
\label{CCFFig}
\end{figure}

In considering a dataset as rich and inhomogeneous as this it is
helpful to divide the data into sub-samples based on behaviour. One
important sub-sample comprises segments of
lightcurves when rapid optical flaring is occurring. The best for this
purpose are June 19 (McDonald night 2), and June 26 (Ultracam night
5), as both provided 1~s time resolution optical lightcurves
spanning relatively extended periods of flaring.  On June 19, we also
have data for a second earlier period of rapid flaring, albeit at
lower time-resolution, from MDM night 1.  We show the June 19 CCFs in
Fig.~\ref{CCFFig}a and b, and the June 26 CCF in Fig.~\ref{CCFFig}c.
The width of a peak in the cross-correlation function indicates the
timescales of correlated variability. A narrow peak is only possible
if both the X-ray and optical variability are dominated by short
timescales; a slow variation in one or both lightcurves will smear the
correlation over a larger range of lags leading to a broader peak. We then
expect that correlations involving the rapid optical flaring will be
narrow and those involving the slower variations will be broad.
The MDM data show that the rapid flaring is the
dominant correlation during the first period of flaring on June 19,
providing a narrow peak near zero lag. This is consistent with
Ultracam flaring studies which suggest a $\sim0.1$~s lag
\citep{Gandhi:2016a,Gandhi:2017a}. The rapid response associated with
fast flaring has now been seen in a number of other BHXRTs as
summarized in Section~\ref{IntroSection}.  During the second flaring
period on June 19, we clearly also see a sharp peak near zero lag, but it is
superposed on a lagged and smeared response. On June 26, the CCF is
dominated by the slow variations, but a small sharp peak can also be
seen close to zero lag. In these latter two cases, then {\em both}
rapid flaring and slow variations are present, correlated with the
X-rays, and seen in the CCF.

We also examine the evolution of the CCF through the large X-ray flare
seen on June 21 in Fig.~\ref{BigFlareCCFFig}. For this night, we have
WHT/Ultracam data from the beginning of the night, followed by MDM
data for the remainder. While the X-ray flare exhibits the largest
X-ray flux seen in our simultaneous dataset, and a dynamic range of
about a factor of fifty, the optical data only vary by a factor of
two, and do not obviously correlate with the X-ray flare. A
correlation on shorter timescales is clearly seen in the Ultracam
lightcurves (Fig.~\ref{UltracamFig}b) and is reflected by a pronounced
CCF peak (Fig.~\ref{BigFlareCCFFig}), but the MDM CCF appears
dominated by a general rise which is likely due to uncorrelated
variability. On closer examination a peak can be seen close to zero
lag; the evolution of this peak is shown in
Fig.~\ref{BigFlareCCFFig}.  Combining these four snapshots of the CCF
evolution, we see that the CCF is quite broad and strongly lagged
($\sim35$~s) before the flare, similar to the lagged and smeared
response seen at other times. During the flare rise, the peak moves to
shorter lags ($\sim 10$~s) and the response becomes narrower, although
with an asymmetric tail to longer lags. Near the peak of the flare a
short episode of rapid flaring occurs (labeled B in
Fig.~\ref{MDMFig}c), during which a very narrow peak occurs, similar
to those discussed above for June 19 and 26, before the CCF reverts to
the short peak delay, extended tail form for the decline from the
flare.

\section{Discussion}
\label{DiscussionSection}

\subsection{Binary geometry and lag expectations}
\label{GeometrySection}

Recurrent X-ray/optical correlations indicate that the optical
emission is causally linked in some way to the central engine around
the black hole, either because one is driving the other, or because
they share a common cause.  They could arise from the same region, for
example the base of the jet, or from different regions coupled by
irradiation or plasmon ejection.  To understand optical variations,
then, we need to understand the binary geometry and the mechanism(s)
coupling X-ray variations to the optical.  We have seen that a lag of
about a minute is common in slower timescale variability, and that
this is accompanied by comparable smearing of the signal. An
  independent analysis by \citet{AlfonsoGarzon:2018a} of individual
  flares also identified a class of optical flares lagging the X-rays
  by $<2$~min). This lagged,
smeared correlation could naturally arise either from light travel
timescales within the binary if the optical is produced by
reprocessing of high energy irradiation \citep{OBrien:2002a}, or from
expanding plasmons in the jet becoming optically thin at progressively
longer wavelengths with time \citep{vanderLaan:1966a}.

Light travel timescales are straightforward to calculate given
estimates for binary parameters. For V404~Cyg, assuming orbital period
$P=6.47$~days, mass ratio $q=0.06$, and black hole mass
$M_1=9.0$~M$_{\odot}$ \citep{Khargharia:2010a}, the binary separation
is about $2.2\times10^{12}$~cm. Assuming the disc is limited to
90~percent of the Roche lobe radius, the maximum disc radius is about
$1.2\times10^{12}$~cm.  Light travel time delays from the disc should
then range from 0--80~s, with the response weighted towards shorter
lags within the range \citep{OBrien:2002a}. Similar estimates
  were made independently by \citet{AlfonsoGarzon:2018a}.

Wavelength dependent lags due to expanding plasmons are harder to
estimate reliably, depending on plasmon
properties. \citet{Rodriguez:2015a} suggested optical lags should be
$\ga10$~min, but this is presumably based on analogy with other
sources. In the specific case of V404~Cyg, we can find ground truth in
the sub-mm to radio range where signatures of this behaviour are
clearly seen with frequency-dependent lags consistent with a plasmon
model \citep{Tetarenko:2017a}. This model predicted that the radio
emission lagged the actual ejection events by tens of minutes. In the
model, the frequency dependent lags, given by their equation 15, scale
approximately as $\nu^{-0.33}$ to $\nu^{-0.5}$ with different indices
corresponding to the range of electron energy distributions inferred
across a range of flares. We can use this lag prescription from
\citet{Tetarenko:2017a} together with their tabulated parameters for
the eight ejections identified to extrapolate the range of lags
expected in the $r$ or $R$ band. The expected lags range from
20--110~s, shorter than expected by \citet{Rodriguez:2015a}.

\subsection{Rapid variability and the central engine geometry}
\label{EngineSection}

The timescale of rapid optical flaring (during periods labelled B in
the lightcurves) is much less than the light-travel timescale across
the disc estimated above, so this behaviour cannot arise in
reprocessing. At the same, time, the flaring does at times clearly
correlate with that seen by JEM-X (see Section~\ref{EchoSection}), so
neither can this arise from local variability (e.g. due to magnetic
reconnection; \citealt{Zweibel:2009a}) in the outer disc. It is most
likely then that this rapid optical flaring originates close to the
central engine producing X-rays. It probably arises in direct jet
emission, although the timescale is much shorter than that estimated
above for plasmon ejection \citep{Gandhi:2016a,Gandhi:2017a},
suggesting a different variability mechanism. Direct optical
  emission from the corona has also been proposed by
  \citet{Dallilar:2017a}.\citet{AlfonsoGarzon:2018a} considered
  similar possibilities for flares with no optical lag.

Based on {\it NuSTAR}, {\it Swift}, and {\it INTEGRAL} observations an
X-ray picture of the central source geometry has emerged, although
many features remain in question. The X-ray spectrum is dominated by
power-law components, presumably associated with inverse Compton
scattering in hot coronal material or the base of the jet, with
superposed Compton reflection \citep{Walton:2017a}. A disc black body
component may have also been present at the lowest energies, although
\citet{Motta:2017b} could not confirm this with {\it Swift}. The
coronal X-ray source illuminating this putative inner disc is required
to be compact, with height less than ten gravitational radii.  Strong
and variable absorption occurred ubiquitously (see
Section~\ref{AbsorptionSection}).  \citet{SanchezFernandez:2017a}
identify the source as predominantly being in the canonical hard state
\citep{Remillard:2006a}, which is usually associated with a recessed
disc. These hard state periods coincide with hard ISGRI
colours. Periods of softer ISGRI colours are identified as transitions
into intermediate or ultra-luminous states by
\citet{SanchezFernandez:2017a} and could then be associated with the
disc moving inwards.  \citet{Walton:2017a} argue that X-ray flares are
likely also associated with transient jet ejections. Synthesizing
these results, a possible scenario is then that the disc is recessed
between large flares, with a quasi-steady compact jet. Transient
instabilities in the jet could give rise to rapid optical
flaring \citep{Malzac:2018a}. During major flares, the disc moves inward with the flare
then ending with disruption of this transient disc and ejection of a
plasmon, leading to a delayed optical flare \citep{Mirabel:1998a}.
Irradiation of the outer disc during flares is also possible, although
may be inhibited by the large amounts of local absorption, discussed
in section~\ref{AbsorptionSection}. We will focus more on the roles of
plasmon ejection and irradiation in mediating slow variability in
Section~\ref{JetSection}, and for the remainder of this Section will
focus on the fastest optical variability.

This rapid optical flaring, sometimes on sub-second timescales,
is notably redder than the slower variations consistent with an
  origin in synchrotron emission from the base of the jet
\citep{Gandhi:2016a,Gandhi:2017a}. Between our June 18 and June 19
McDonald observations, when several of these rapid flaring episodes
occurred, \citet{Trushkin:2015a} reported a radio detection with an
inverted spectrum, suggesting optically thick emission from a compact
jet, adding support to the idea that jet activity was present at this
time.  Based on an 0.1~s delay between X-ray and rapid optical flares,
the optical emission region producing these rapid variations is
inferred to lie $\la0.1$~light seconds (about 1000 Schwarzschild
radii) above the black hole, and so is part of the central engine
region.

We find that this rapid optical flaring is associated with the
lower-envelope in the X-ray/optical flux relationship
(Section~\ref{FluxCorrelationSection}). Times when rapid flaring are
present are associated with times when the hard X-ray colour (the
ISGRI colour) is quite hard (Section~\ref{XrayColourSection}). 
  As noted above, hard ISGRI colours are identified by
  \citet{SanchezFernandez:2017a} with the canonical hard state. Our
  lower envelope then appears to correspond to this hard state, as do
  rapid optical flares. This is consistent with behaviour in other
  BHXRTs.

The rapid optical flares themselves can quite suddenly
appear and disappear with no change in X-ray intensity or colour
(e.g. Fig.~\ref{FlareTransitionFig}) and they do not appear to be
immediate consequences of changes in the X-ray state.  These changes
in optical flaring may represent 
variations within the jet itself
downstream from where X-rays are produced, such that sometimes
intense optical variability is seen and sometimes it is not. One
proposed mechanism for this rapid variability is internal shocks
within the jet caused by fluctuations in the jet velocity
\citep{Malzac:2018a}.  

\subsection{Are slow variations driven by plasmon ejections or disc irradiation?}
\label{JetSection}

Rapid optical variability in V404~Cyg (second-timescale and
  faster) is clearly associated with the central engine and jet, as
  discussed above.  This rapid flaring is confined to the lower
  envelope of the X-ray vs. optical relationship, but except for the
  peaks of extremely transient flaring, it never dominates the optical
  flux.  It remains to elucidate then whether this correlation itself
  reflects jet behaviour, with the slower optical variations that
  dominate it arising in expanding plasmons downstream of the jet
  base, or whether the correlation instead reflects irradiation of a
  disc or wind component. Both effects are expected to be present at
  some level, and both produce a correlation comparable to
  $F_{\rm opt}\propto F_{\rm X}^{0.6}$ \citep{Russell:2006a}, and so
  jet and disc may both be contributing to the behaviour.

  As noted in Section~\ref{GeometrySection}, radio and sub-mm
  observations indicate that larger scale plasmon ejections did occur
  at other times in this outburst. In particular,
  \citet{MillerJones:2019a} actually resolve radio ejections on
  several occasions, with most intensive coverage on 2015 June 22 (MJD
  57,195. \citet{Tetarenko:2017a} show
  multi-frequency lightcurves with simultaneous X-ray and optical
  coverage through a series of flares also on June 22,
  beginning with the large flare caught at the end of the fourth night
  of our MDM data. Three X-ray flares are covered, two large and one
  small. The ends of the two large ones both show an optical decay
  similar to our exponential decays, and both are followed by flares
  at sub-mm, then mm, then radio frequencies.  A model of expanding
  plasmons \citep{vanderLaan:1966a} provides remarkable agreement with
  the multi-frequency lightcurves and the set of flares observed can
  be well modelled by assuming a series of eight ejection events, with
  the X-ray cut-off and optical decay corresponding to two of the
  events.

The optical behaviours we saw on June 19, and that reported on June 22
by
\citet{Tetarenko:2017a}, both resemble that seen in GRS~1915+105, where a
disappearance of X-ray flux is also followed by an infrared, and then
a radio flare \citep{Mirabel:1998a}. Those events were interpreted as
due to disruption of the inner disc and ejection of an expanding
plasmon, which then emits at progressively longer wavelengths as
described above.  We see optical lags of order a minute in V404~Cyg,
in contrast to the 15~minute IR lag seen in GRS~1915+105
\citep[c.f.][for claimed longer
lags]{Rodriguez:2015a,AlfonsoGarzon:2018a}. As estimated above,
however our shorter optical lag, around one minute, is expected from
the ejection model of \citet{Tetarenko:2017a} based on the flares they
observe in V404~Cyg.  A one minute lag would also be consistent with
light travel timescales across the disc, so it remains possible that
either the slow optical variations arise from direct emission from the
plasmons, or from irradiation of the disc or wind by the X-ray flares,
or by a combination of both mechanisms. \citet{Dallilar:2017a} have
instead argued that decay events similar to those shown in
Section~\ref{DecaySection} occur in stationary material in the corona
or base of the jet with the optical decay reflecting synchrotron
cooling with no change in magnetic field, rather than the expansion of
ejecta.

While the overall lag may be comparable between different models,
  the wavelength dependence of the lag may differ as plasmon models
  predict a strong frequency dependence.  The high time-resolution
  multicolour observations of \citet{Gandhi:2016a} show lags between
  optical bands, $\sim4$~s. This is in reasonable agreement with the
  $\la 10$~s inter-band lags that would be expected based on the
  \citet{Tetarenko:2017a} ejection model
  (Section~\ref{GeometrySection}.), but does not rule out irradiation
  which also predicts shorter lags at shorter wavelengths.

  Optical colour variations during flares can also be helpful to
  disentangle variability mechanisms.  \citet{Gandhi:2016a} find that
  rapid flaring is quite red, as expected for optically thin
  synchrotron emission, while the slower variations are bluer. The
  latter could be consistent with either disc reprocessing or
  self-absorbed synchrotron.  \citet{Tachibana:2017a} argued that the
  slowly variable component could be further decomposed into a
  `highly-variable component' with blue colours and flux rising with
  frequency, and a redder `little variable component' with
  approximately flat colours invoking both jet and disc emission to
  explain the components.  \citet{Kimura:2016a} argue based on the
  spectral energy distribution that the primary optical response to
  X-rays is mediated by reprocessing while \citet{Maitra:2017a}
  suggest that the stability of the colours through large flux changes
  is inconsistent with irradiation which should become bluer as it
  brightens.

  If large optical flares do arise in the jet, then we would also
  expect to see polarization signatures. While an optically thick jet
  would be free of polarization, by its nature the expanding plasmon
  model produces flare peaks when the plasmon transitions from
  optically thick to thin at the observed wavelengths, so polarization
  {\em should} be seen at least during the decay of flares when
  emission is optically thin. Infrared polarization has now been
  detected from several quiescent, outbursting, and persistently
  accreting black hole binaries, lending credence to this expectation
  \citep{Russell:2008a,Russell:2014a,Russell:2016a,Chaty:2011a}. Unfortunately,
  here too the observational evidence is equivocal with some studies
  finding variable polarization, and others not, and characteristics
  that may be more consistent with scattering in a disc outflow rather
  than from jet synchrotron emission
  \citep{Tanaka:2016a,Lipunov:2016a,Shahbaz:2016a,Itoh:2017a,Kosenkov:2017a}.

In summary, both jet emission, and X-ray irradiation remain plausible
mechanisms to couple the slow variations in the optical, with evidence
in favour of both interpretations from colour changes,
  inter-band lags, and polarization.  It is possible that both
  mechanisms do
play a role, and that the inconsistent and equivocal evidence
discriminating them is a consequence of this
  \citet[c.f.][]{AlfonsoGarzon:2018a}. The apparent bimodal
behaviour seen in Fig.~\ref{FluxCorrFig} also suggests two distinct
modes of coupling the X-ray to the optical, with the lower envelope
showing characteristics typical of optical jet emission such as the
$F_{\nu} \propto \nu^{0.6}$ dependence, rapid optical flaring and the
association with hard X-ray colour. More speculatively, the
upper branch might represent a disc dominated mode with the optical
coupled to X-rays at least partly through irradiation.

\subsection{Local absorption}
\label{AbsorptionSection}

There is considerable evidence for large intrinsic absorption from the
inner accretion disc 
  \citep{Motta:2017a,SanchezFernandez:2017a}. The soft
  hardness-intensity diagram we show in Fig.~\ref{XColFig}b also shows
  behaviour suggesting variable X-ray absorption with a track where
  the source fades and becomes harder. Only a subset of data follow
  this track, however, and another group of points suggest softening
  as the source fades. Absorption then may account for some of the
  variability we see, but not all of it.

Absorbing material may impact not only our view of X-rays, but
  also the ability of X-rays to irradiate the outer disc and wind. It
  also means that the X-ray lightcurve we see may not be the same as
  the outer disc sees. This could partially explain the
  inconsistencies that we see between X-ray and optical flare
  amplitudes.  \citet{Motta:2017b} suggest that X-ray behaviour
  implies two layers of absorption: a high column density
  inhomogeneous region within 100 gravitational radii, and a more
  diffuse and homogeneous one at larger radii. The inhomogeneous
  absorption then occurs inside the optically dominant part of the
  disc. Strong optical flares with weak X-ray counterparts could
arise if the X-ray emission is obscured but optical emission is
visible.  Absorption will affect high energy photons least, so the
ISGRI lightcurve, which usually does resemble the optical one more
closely than the JEM-X lightcurve does, should better trace the
central activity.  In an irradiation model the optically emitting
regions of the disc will be directly visible, and are likely to be
illuminated by many lines of sight at different azimuths, so they will
not see the same localized absorption as we see in X-rays. In a jet
model, optical synchrotron flares associated with transient jets are
expected to be emitted from beyond $\sim1000$ Schwarzschild radii
above the black hole \citep{Gandhi:2017a}.  The inner absorption
region inferred by \citet{Motta:2017b}is then likely too small to
obscure the jet base, and the outer region too diffuse and
homogeneous.

Obscuration of direct X-ray emission does not explain cases like
Fig.~\ref{OptFlareFig} where several X-ray flares are seen, with only
the last clearly echoed in the optical; it is hard to obscure
only the optical emission. If the optical emission is associated
with irradiation, there could be a low-altitude obscuring torus that
can completely impede irradiation of the outer disc, while we can see
over it. Only when this torus mostly dissipates (possibly at the end
of a series of flares as in Fig.~\ref{OptFlareFig}), is the optical
disc exposed to irradiation. This kind of behaviour may also
  explain the cases identified by \citet{AlfonsoGarzon:2018a} where
  large optical lags are seen.

\subsection{Winds}
\label{WindSection}

Optical spectroscopic monitoring revealed a strong, high velocity, but
low-excitation wind from the outer disc which appeared to carry away
most of the pre-outburst disc mass and terminate the outburst
  around the time of our last time-resolved observations
  \citep{MunozDarias:2016a,Rahoui:2017a}.  The strong wind from the
outer disc inferred by \citet{MunozDarias:2016a} will have a profound
impact on the kind of observations we are making. We can place a lower
limit on the density of such a wind quite
readily. \citet{MunozDarias:2016a} estimate that a total mass of at
least $10^{-8}{\rm M}_{\odot}$ is expelled over $\sim10$~days of
outburst, implying an average mass loss rate
$\ga2\times10^{19}{\rm g}{\rm s}^{-1}$. If this emerges uniformly over
the whole disc (radius $1.2\times10^{12}{\rm cm}$) at $0.01{\rm c}$,
then the density of the wind should be
$\sim1.5\times10^{-14}{\rm g cm}^{-3}$, corresponding to a particle
density of $9\times10^{9}$~cm$^{-3}$ for pure hydrogen, comparable to
density estimates made by \citet{King:2015b} based on X-ray line
emission measures.  Finally, if the central source is viewed near
edge-on through this wind (as is the case in irradiating the outer
disc), then we expect a column of $0.02{\rm g cm}^{-2}$, or
$N_{\rm H}\sim 1\times10^{22} {\rm cm}^{-2}$, again comparable to
X-ray based estimates given by \citet{King:2015b}. The uncertainties
in this estimate: adopting the minimum mass loss suggested by
\citet{MunozDarias:2016a}, assuming the whole area of the disc
participates in the wind and using the terminal velocity of an
accelerating wind all tend to lower the inferred column density, so
this can be considered a lower limit on the {\em mean} column. If the
mass loss is episodic, as is likely given the extremely variable
irradiation of the disc, then at times it will be lower, but at other
times it may be much higher. This minimum mean column is about twice
the interstellar column. It will produce an enhanced absorption in
spectroscopy, but should not completely block our view of the central
source and neither will it prevent irradiation of the outer disc,
unless the column is episodically or locally much higher.

If large optical variations do arise from reprocessing then the
presence of such a strong wind as is inferred will modify both the
irradiation itself, and the geometry of the reprocessing
material. Signatures of irradiation may then be quite different to
those in more normal LMXBs.  One effect that can be expected from such
a wind is that soft X-rays will only penetrate partially into the
wind, leaving the outer region relatively shielded from ionizing
radiation.  With this picture, such soft X-rays as are present in an
already hard spectrum will be absorbed by the wind, leaving only the
harder X-rays to irradiate the outer disc.  These will penetrate deep
into the photosphere, leading to quite long reprocessing times
\citep{Cominsky:1987a,McGowan:2003a} which may substantially enhance
the observed lags compared to light travel times alone.
%Furthermore,
%if the photosphere is itself expanding as the base of the wind, as
%expected from the thermally driven wind inferred, then the scale
%height will be increased compared to a disc in hydrostatic
%equilibrium, and photons will travel further in diffusing out of the
%disc, further increasing reprocessing times.
We then expect that the
disc reprocessing signal will be lagged by a combination of
light-travel times ($\la 80{\rm s}$), and significant reprocessing
times. Observing lags longer than possible from geometrical light
travel times alone is then not a problem for the reprocessing model,
and these prolonged reprocessing times could contribute to the
extended tails to smooth decays described in
Section~\ref{DecaySection}. In this respect we disagree from
  \citet{AlfonsoGarzon:2018a} who only consider geometrical light
  travel time delays for reprocessing.

\section{Conclusions}

We have performed a serendipitous campaign monitoring the relationship
between optical and X-ray variability in V404~Cyg during the rise and
peak of the outburst of 2015 June.  We focus on variability on
timescales of seconds and longer, finding large amplitude optical
variability, usually related to X-ray behaviour, although not
perfectly correlated.

We find a relationship between X-ray and optical fluxes bounded by
lower and upper envelopes. These envelopes converge around the
Eddington limit.  The lower envelope is traced by a power-law relation
$L_{\rm opt}\propto L_{\rm X}^{0.54}$.  At times we see motion along
this line within a night, reinforcing its significance as a boundary
on the source behaviour. This lower envelope is consistent in
power-law index with X-ray/optical relations identified previously by
other authors and can approximately be extrapolated all the way to
quiescent fluxes.  We find a correlation between the position in the
X-ray/optical diagram and the X-ray spectrum, with locations close to
the lower-envelope corresponding to harder hard X-ray colours
([60--200~keV] / [25--60~keV]). These hard X-ray hard states are
identified by \citet{SanchezFernandez:2017a} as the canonical hard
state. A subset of data seem to lie above this relationship and
distinct from it, defining a an upper envelope approximately tracing a
$L_{\rm opt}\propto L_{\rm X}^{0.12}$ relation. These periods are
identified by \citet{SanchezFernandez:2017a} as likely corresponding
to intermediate and ultra-luminous states in other sources. As well as
the X-ray/optical flux correlation, we find the hard X-ray hardness is
anti-correlated with the optical brightness and positively correlated
with R.M.S. optical variability. It is likely that all of these
properties (X-ray and optical flux, hard X-ray colour, and optical
rapid variability) are driven by changes in the central accretion rate
and geometry. Such changes will directly affect the X-ray spectrum,
and optical emission from the jet. They may also affect the efficiency
with which the disc can reprocess X-rays, and so affect the slowly
varying component of optical emission as well. 

The optical variability is quite diverse, but can broadly be divided
into two types.  Contributing most of the variability are large
amplitude, slow variations that probably arise either in X-ray
irradiation of the outer disc and its wind, or from expanding jet
ejecta, or both. In this respect, our conclusions agree with the
  recent work of \citet{AlfonsoGarzon:2018a} which was based on quite
  different analysis focused on identifying and classifying individual
  flaring events. These slow variations generally follow X-ray, and
especially hard X-ray, behaviour, with a lag and smearing on
$\sim$minute timescales. The connection between X-ray and optical is
complex, with at times a strong optical response to a weak X-ray flare
or vice versa. In part, this is probably due to variable obscuration
of the central X-ray source either from our perspective or from that
of the outer disc.  During extended X-ray quiet periods we see a
prolonged tail to this slow response that is seen to continue to fade
for up to 1000~s. This is much longer than light travel times in the
binary, so if they arise from reprocessing then they require quite a
large diffusive reprocessing time to release the energy absorbed from
irradiation.  The other, and apparently distinct, component of
variability is transient rapid flaring showing small or no lag with
respect to X-rays. These flares occur simultaneously with the slower,
lagged variations and both can then be seen simultaneously in
cross correlation functions. The rapid flares appear to be direct
synchrotron emission from a standing shock at the base of the jet
\citep{Gandhi:2017a}. These only occur near the lower envelope of the
X-ray/optical flux relationship and appear analogous to rapid optical
flaring observed in other hard state sources as summarized in
Section~\ref{IntroSection}. Sometimes these periods are very
short-lived, with a single very large flare. At other times the
optical flaring can last for periods $\sim10$~minutes, then fade away
with no apparent change in X-ray colours or variability. Since the
rapid optical flaring appears to be associated with the central source
itself, this may indicate intrinsic changes in optical production from
the base of the jet, rather than effects of absorption.

The 2015 activity of V404~Cyg has provided a fascinating, if short,
view of accretion onto BHXRTs. In some respects, including the X-ray
state evolution and presence of rapid optical synchrotron flaring it
follows patterns identified in other sources. The large orbital period
of the system means that the accretion disc is larger than in almost
other systems, except for GRS~1915+105 which has its own
peculiarities, and this may be responsible for the unusual outburst
evolution and strong mass loss which played a large role in the
outburst. The impact of outflows on our observations is magnified by a
relatively high inclination which ensured our view of the central
source was heavily impacted by absorption or complete obscuration by
material above the disc. It appears, then, that while the outburst
showed many unusual and unprecedented features, these owe more to the
large scale environment and viewing angle than to the behaviour of the
central source itself.

\section*{Acknowledgements}

We thank John Morales for assistance with the McDonald Observatory
observations and an anonymous referee for many suggestions which
improved this paper. We are deeply grateful to the {\it INTEGRAL}
mission for providing intensive coverage of this outburst and
committing to provide prompt public access to both raw and processed
data. We are also grateful to Christian Knigge for coordinating
collaborative efforts on this source, without which the multiple
datasets included here likely would not have been combined in this
way.

This work is based in part on observations obtained at the MDM
Observatory, operated by Dartmouth College, Columbia University, Ohio
State University, Ohio University, and the University of Michigan.
D.~M.~T. acknowledges support from the National Science Foundation,
via grant AST-1411685 to The Ohio State University. P.~G. acknowledges
support from STFC (grant ST/R000506/1). V.~S.~D., T.~R.~M. and
ULTRACAM are also supported by STFC.

This paper makes use of data obtained as part of the INT Photometric
H-alpha Survey of the Northern Galactic Plane (IPHAS, www.iphas.org)
carried out at the Isaac Newton Telescope (INT). The INT is operated
on the island of La Palma by the Isaac Newton Group in the Spanish
Observatorio del Roque de los Muchachos of the Instituto de
Astrofisica de Canarias. All IPHAS data are processed by the Cambridge
Astronomical Survey Unit, at the Institute of Astronomy in
Cambridge. The bandmerged DR2 catalogue was assembled at the Centre
for Astrophysics Research, University of Hertfordshire, supported by
STFC grant ST/J001333/1.

This research has made use of data obtained through the High Energy
Astrophysics Science Archive Research Center Online Service, provided
by the NASA/Goddard Space Flight Center.

%%%%%%%%%%%%%%%%%%%%%%%%%%%%%%%%%%%%%%%%%%%%%%%%%%



%\bibliographystyle{mnras}


%%%%%%%%%%%%%%%%%%%%%%%%%%%%%%%%%%%%%%%%%%%%%%%%%%

% Don't change these lines
\bsp	% typesetting comment
\label{lastpage}
\end{document}